\documentclass[usenatbib,preprint]{mn2e}
\usepackage{graphicx}
\usepackage {natbib,aas_macros}
\usepackage {bm}
\usepackage {amsmath,amssymb}
\usepackage {setspace}
\usepackage{fixltx2e}

\newcommand{\cs}{c_{\rm s}}

\newcommand{\Ml}{M_{\rm L}}
\newcommand{\deldel}[2]{\frac{\partial #1}{\partial #2}}
\newcommand{\epf}{\Omega_{\rm epi}}
\title[Condition for disk fragmentation]{A revised condition for self-gravitational fragmentation of protoplanetary disks}

\author[S. Z. Takahashi et al.]{S. Z. Takahashi$^{1,2,3}$, Y. Tsukamoto$^{4}$,
and  S. Inutsuka$^{3}$ \\
$^1$Astronomical Institute, Tohoku University, 6-3 Aoba, Aramaki-aza, Aoba-ku, Sendai, Miyagi,
Japan\\
$^2$Department of Physics, Kyoto University, Oiwake-cho,
Kitashirakawa, Sakyo-ku, Kyoto, Japan \\
$^3$Department of Physics, Nagoya University, Furo-cho, Chikusa-ku,
Nagoya, Aichi, Japan  \\
$^4$RIKEN, Wako, Saitama, Japan\\
}

\begin{document}
\maketitle
\begin{abstract}
Fragmentation of protoplanetary disks due to gravitational instabilities is a candidate of a formation mechanism of binary stars, brown dwarfs, and gaseous giant planets. 
The condition for the fragmentation has been thought that the disk cooling timescale is comparable to its dynamical timescale. 
However, some numerical simulations suggest that the fragmentation does not occur even if the cooling time is small enough, or the fragmentation can occur even when the cooling is inefficient. 
To reveal a realistic condition for fragmentation of self-gravitating disks, we perform two-dimensional numerical simulations that take into account the effect of the irradiation of the central star and radiation cooling of the disk, and precisely investigate the structure of the spiral arms formed in the protoplanetary disks. 
We show that the Toomre $Q$ parameter in the spiral arms is an essential parameter for fragmentation. 
The spiral arms fragment only when $Q <0.6$ in the spiral arms.
We have further confirmed that this fragmentation condition observed in the numerical simulations can be obtained from the linear stability analysis for the self-gravitating spiral arms. 
These results indicate that the process of fragmentation of protoplanetary disks is divided into two stages: formation of the spiral arms in the disks; and fragmentation of the spiral arm. 
Our work reduces the condition for the fragmentation of the protoplanetary disks to the condition of the formation of the spiral arm that satisfies $Q < 0.6$.
\end{abstract}


\section{Introduction}
\label{intro}
The formation of stars requires gravitational collapse of dense molecular cloud cores.
The run-away collapse of the central dense region in the self-gravitationally collapsing core results in the formation of the first hydrostatic core of which mass is of the order of $10^{-2} M_\odot$ \citep{1969MNRAS.145..271L,1998ApJ...495..346M}. 
The subsequent dynamical collapse of the first core triggered by the dissociation of molecular hydrogen proceed in a run-away manner again and result in the protostar of which mass is of the order of $10^{-3} M_\odot$ \citep[e.g.,][]{2000ApJ...531..350M}. 
The evolution after the emergence of the protostar is called the {\it accretion phase}. 
Since the parental molecular cloud cores have angular momentum \cite[]{1993ApJ...406..528G, 2002ApJ...572..238C}, the infalling material in the cloud cores cannot accrete directly onto the protostars, and thus, the rapidly rotating circumstellar disks are formed around the protostars.
The circumstellar disks are also called protoplanetary disks since they are supposed to be the sites of planet formation. 
Numerical simulations of formation of protostars and circumstellar disks suggest that the disks are very massive in its early evolutionary phase \cite[cf.][]{1998ApJ...508L..95B, 2010ApJ...718L..58I, 2011MNRAS.416..591T}.
In the massive disk, spiral arms are formed by the gravitational instability \cite[]{1976PThPh..56.1665T, 1978PASJ...30..253T, 1978PASJ...30..223I}. 
The angular momentum of the protoplanetary disks is redistributed by the gravitational torque of the spiral arms and the angular momentum transfer promotes gas accretion in the protoplanetary disks.

When the disks are massive enough, the spiral arms become gravitationally unstable and fragmentation occurs \cite[e.g.][]{2003ApJ...595..913M, 2010ApJ...724.1006M, 2011ApJ...729...42M, 2010ApJ...708.1585K, 2010ApJ...714L.133V,2011MNRAS.416..591T, 2012PASJ...64..116K, 2012ApJ...746..110Z,2013MNRAS.436.1667T}.
Recent observations have revealed that the gaseous giant planets exist in wide orbits of more than 30AU \cite[cf.][]{2008Sci...322.1348M,2010Natur.468.1080M, 2009ApJ...707L.123T, 2009A&A...493L..21L,2010ApJ...719..497L, 2013ApJ...763L..32C}.
Since it is difficult to form such wide orbit gaseous giants through the classical core accretion scenario \cite[]{1985prpl.conf.1100H}, the fragmentation of the disks due to gravitational instability is the most promising scenario for forming such planets.
The self-gravitational fragmentation of the circumstellar disks is also important as a formation process of multiple systems \cite[]{2008ApJ...677..327M, 2010ApJ...708.1585K} and brown dwarfs \cite[]{2009MNRAS.392..413S, 2011ApJ...730...32S, 2012ApJ...750...30B}.
Moreover, the fragments formation affects the evolution of the protoplanetary disks.
For example, \cite{2006ApJ...650..956V} and \cite{2011ApJ...729...42M} suggest that the episodic accretion occurs when the fragments are created in the protoplanetary disk. 
They argued that the episodic accretion may be related to FU-Ori-type outburst phenomena.

Despite its importance in the formation of stars and planets, the criterion of the gravitational fragmentation is still controversial. 
The gravitational stability is characterized by Toomre $Q$ parameter:
\begin{equation}
 Q\equiv \frac{\cs\epf}{\pi G \Sigma},
\end{equation}
where $\cs$ is the sound speed, $\epf$ is the epicyclic frequency, and $\Sigma$ is the surface density of the disk.
When the gravitationally unstable disks are formed and satisfy $Q\sim 1$, the spiral arms are developed in the disks.
Then the disks are heated by the shock of the spiral arm, and the surface density is redistributed by the angular momentum transfer due to the gravitational torque. 
These effects make the disks gravitationally stable.
As a result, the self-gravitating disks are self-regulated to the  marginally stable $Q \sim 1$ condition.
Thus we need physical mechanisms that make the disk unstable overcoming the stabilization.

\cite{2001ApJ...553..174G} firstly showed that the fragmentation can happen when the disk cooling timescale is comparable to its dynamical timescale with two-dimensional local shearing box simulation.
This is so-called the cooling criterion.
In \cite{2001ApJ...553..174G}, the cooling rate is modeled with the cooling time $\tau_{\rm c}$, and the criterion of the fragmentation is that the normalized cooling time $\beta\equiv \tau_{\rm c}/\Omega$ satisfies $\beta \lesssim 3$.
\cite{2005MNRAS.364L..56R} have shown that the cooling criterion given by \cite{2001ApJ...553..174G} is consistent with the results of the three-dimensional global simulations.
\cite{2011MNRAS.411L...1M} have questioned, however, that the numerical convergence of $\beta_{\rm crit}$ is not reached in previous studies.
\cite{2011MNRAS.413.2735L} have suggested that the convergence of $\beta_{\rm crit}$ is strongly affected by the heating due to the artificial viscosity.
The value of $\beta_{\rm crit}$ depends on the artificial viscosity:
\cite{2014MNRAS.438.1593R} and \cite{2012MNRAS.427.2022M} have indicated that $\beta_{\rm crit }$ converges to $\sim 8$ and $\sim 30$, respectively.
Because the cooling criterion is simple and useful, many theoretical works on the evolution of the protoplanetary disks have used this condition \cite[e.g.][]{2005ApJ...621L..69R}.

However, there are the several papers that seem to contradict the fragmentation criterion. 
For example, \cite{2010ApJ...718L..58I, 2010ApJ...724.1006M} and \cite{2011ApJ...729...42M} show that the fragmentation can occur even with very stiff EOS. 
On the other hand, \cite{2010ApJ...708.1585K} shows that the fragmentation does not occur with isothermal EOS, which corresponds to very small cooling time, and \cite{2015MNRAS.446.1175T} shows the cases in which the fragmentation does not occur even if the cooling time is small enough. 
Moreover, some previous studies have claimed that there is no universal $\beta_{\rm crit}$. 
They have indicated that the condition of the fragmentation depends on the disk mass \cite[]{2008ApJ...681..375K,2010ApJ...708.1585K} and external irradiation \cite[]{2010ApJ...708.1585K}, and that fragmentation have a stochastic nature \cite[]{2012MNRAS.421.3286P,2015MNRAS.451.3987Y,2016MNRAS.455.1438Y}.
Therefore, the simple cooling criterion for the disk fragmentation is not always reliable in the realistic situations.

\begin{figure}
\includegraphics[width=80mm]{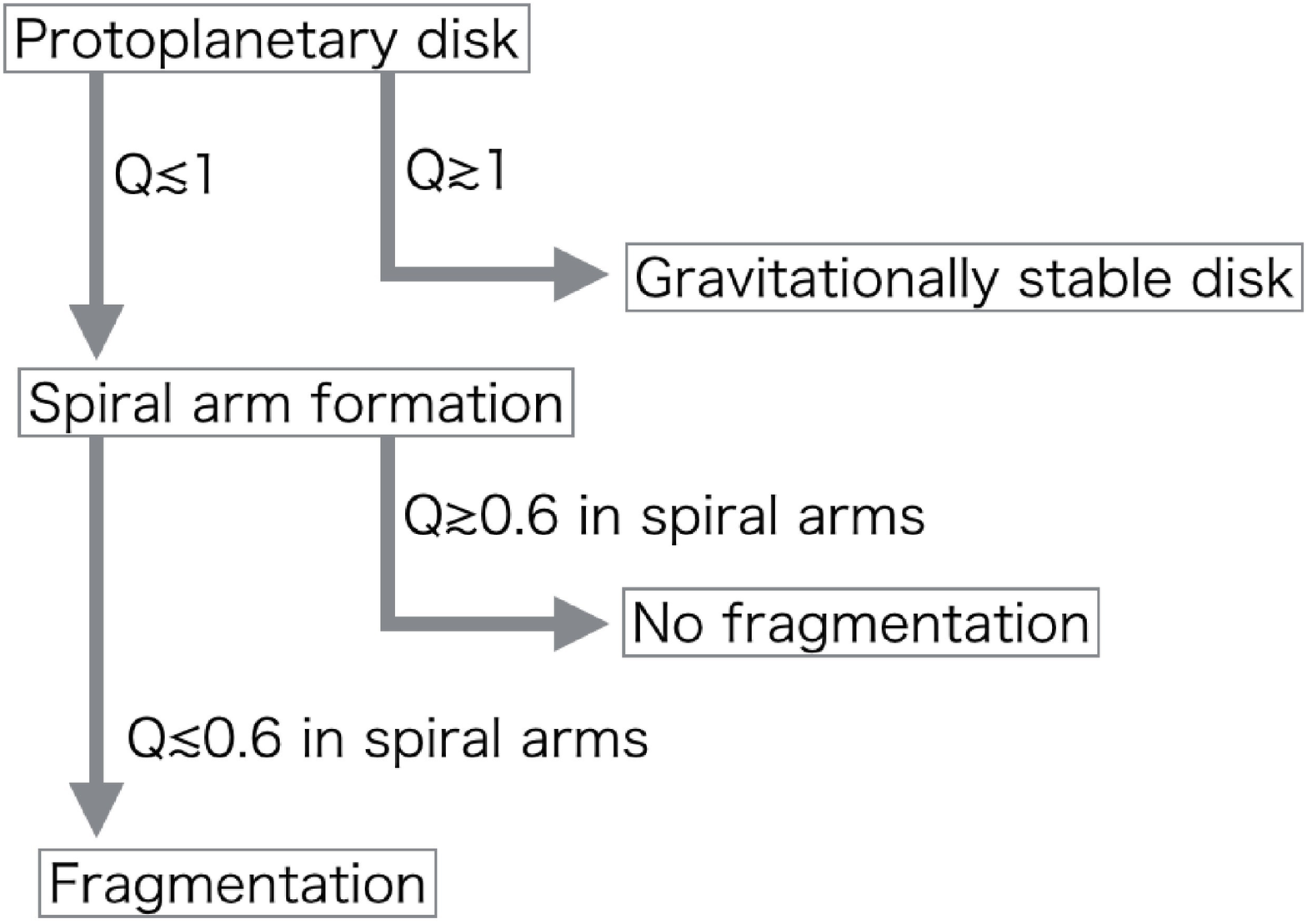}
\caption{Flowchart of fragmentation process of disk found in this work.}
\label{fig:flowchart}
\end{figure}

To find a reliable condition for disk fragmentation, we have to investigate the physical process of the fragmentation precisely.
For this purpose, we divide the process of self-gravitational fragmentation of disks into two stages: the formation of spiral arms in the disk and the fragmentation of the spiral arms.
The nature of spiral arms formed in non-liner growth phase of GI disk depends on numerous factors such as the temperature, mass, and radius of the disk. 
Thus, it seems to be highly difficult to predict the nature of spiral arms formed in disk by few parameters.
On the other hand, as we describe bellow, the fragmentation condition of the spiral arms is surprisingly simple.
In this work, we show that fragmentation of the spiral arms is described by ``gravitational instability of spiral arm''.
We find that the condition for the gravitational instability  of the spiral arms is given by $Q < 0.6 $ in the spiral arms.
Figure \ref{fig:flowchart} shows the flowchart of the fragmentation process of the disk.
In this paper, we mainly discuss the fragmentation process of the spiral arms.

We perform two-dimensional numerical simulations taking into account the effect of the external radiative heating and local radiative cooling.
We confirm that the normalized cooling time $\beta$ cannot predict whether a disk fragments or not in our simulation and 
show that fragmentation of a disk is closely related to $Q$ parameter in spiral arms.
This paper is organized as follows.
In Section \ref{method}, we describe the numerical method.
In Section \ref{result}, we show the results of our simulations and propose the relation between the gravitational fragmentation and minimum $Q$ parameter in the disk.
We confirm the relation found in the numerical simulations by the linear stability analysis of the self-gravitating spiral arms in Section \ref{linear_ana}.
In Section \ref{discussion_frag}, we compare our result with previous papers and discuss the formation processes of the fragments.
We also discuss the relation between the accretion of the gas in the disks and the condition for the fragmentation.
We make conclusions of this work in Section \ref{conclusion_frag}.

\section{Method}
\label{method}
We use the grid-based two-dimensional hydrodynamical simulation code, FARGO-ADSG \cite[]{2000A&AS..141..165M,2008ApJ...678..483B,2008PhDT.......292B}.
\subsection{Basic equations}
We solve the two-dimensional hydrodynamic equations including self-gravity:
\begin{equation}
\deldel{\Sigma}{t} + \nabla\cdot(\Sigma \bm{v}) = 0,
\end{equation}
\begin{equation}
 \Sigma\left(\deldel{\bm{v}}{t}+\bm{v}\cdot\nabla\bm{v}\right)
 = -\nabla  P -\Sigma \nabla \Phi,
\end{equation}
\begin{equation}
 \deldel{E}{t}+\nabla\cdot(E\bm{v}) = -P\nabla \cdot\bm{v} 
  - \Lambda_{\rm C},
\end{equation}
where $\Sigma$ is surface density, $\bm{v}$ is velocity, 
$E$ is internal energy per unit area, 
$P$ is the vertically integrated pressure,
$\Phi$ is the gravitational potential,
$\Lambda_{\rm C}$ is the cooling rate per unit area.
We calculate $\Phi$ with the thin disk approximation.
We assume an ideal gas equation of state,
\begin{equation}
 P = (\gamma -1)E,
\end{equation}
where $\gamma$ is the ratio of specific heat.
We adopt $\gamma = 5/3$ in this calculation.
The temperature is given by 
\begin{equation}
 T = \frac{\mu m_{\rm H}}{k_{\rm B}}\frac{P}{\Sigma},
\end{equation}
where $\mu$ is the mean molecular weight, $k_{\rm B}$ is the Boltzmann constant and $m_{\rm H}$ is the hydrogen mass.
Here we adopt $\mu = 2.34$.
The cooling rate $\Lambda_{\rm C}$ is modeled as follows \cite[]{1990ApJ...351..632H};
\begin{equation}
 \Lambda_{\rm C} = \frac{8}{3} \sigma (T^4 - T_{\rm
  ext}^4)\frac{\tau}{\frac{1}{4}\tau^2+ \frac{1}{\sqrt{3}}\tau + \frac{2}{3}},
\end{equation}
where $\sigma$ is the Stefan-Boltzmann constant, $T_{\rm ext}$ is the equilibrium temperature under the irradiation from the central star, and $\tau = \kappa_{\rm R}\Sigma$ is the optical depth of the disk.
In this work, we do not adopt the cooling function given by constant $\beta$ to investigate the gravitational fragmentation with the realistic cooling process \cite[cf. ][]{2003ApJ...597..131J}.
The Rosseland mean opacity $\kappa_{\rm R}$ is given by 
\begin{equation}
 \kappa_{\rm R} = \kappa_{10} \left(\frac{T}{10 [\rm K]}\right)^2  
{\rm  cm^2 g^{-1}}.
\label{eq:opacity}
\end{equation}
This modeling approximates to the result of \cite{2003A&A...410..611S} in $T\lesssim$ 200 K.
In our simulations, the  temperature is smaller than 200 K except for fragments.
We assume $T_{\rm ext}$ as follows \cite[cf.][]{1997ApJ...490..368C};
\begin{equation}
 T_{\rm ext} = \max\left[
		T_{100}  \left(\frac{r}{100 {\rm au}}\right)^{-3/7},
		\ 10 {\rm K} 
		\right],\label{eq:T_ext}
\end{equation}
where we adopt $T_{100}=20.8 \ {\rm or}\  83.4$K as the temperature at $r=100$ au. 

\subsection{Numerical procedure and Initial conditions}
We use two-dimensional polar grid $(r, \phi)$.
The inner and outer boundaries are  $r= R_{\rm in} \ {\rm and }\ 1000$ au, respectively.
Open boundary conditions are used for the inner and outer boundaries.
The radial and azimuthal grid numbers are 512 and 1024, respectively, and the radial spacing is logarithmic. 
The mass of central star is $M_*=0.5M_{\rm \odot}$ and we take into account the motion of the central star by including the indirect potential. 
The radial and azimuthal components of the self-gravity are smoothed over a softening length. 
We adopt a softening length 0.012$r$ or 0.03$r$, which is larger than the cell size so as to reach the numerical convergence \cite[cf.][]{2015MNRAS.451.3987Y}. 
The influence of the softening length is discussed in Section \ref{softening_length}.

Initial surface density and temperature profiles are given by
\begin{equation}
 \Sigma = \Sigma_{100}\left(\frac{r}{100 {\rm au}}\right)^{-12/7}
  \exp\left(-\frac{r}{r_{\rm disk}}\right),
\end{equation} 
\begin{equation}
 T = 4T_{100}\left(\frac{r}{100 {\rm au}}\right)^{-3/7}.
\end{equation}
where $r_{\rm disk}$ is the outer edge of the disk.
We adopt here $r_{\rm disk} = 250$ au.
With these initial conditions, the initial value of $Q$ is almost uniform for  $r<r_{\rm disk}$.
Initial radial velocity is set to be zero, and initial rotational velocity is given so as to satisfy the force balance in the radial direction. 
The parameters explored in our simulations are summarized in Table \ref{tab:models}.

\begin{figure}
\begin{center}
 \includegraphics[width=80mm]{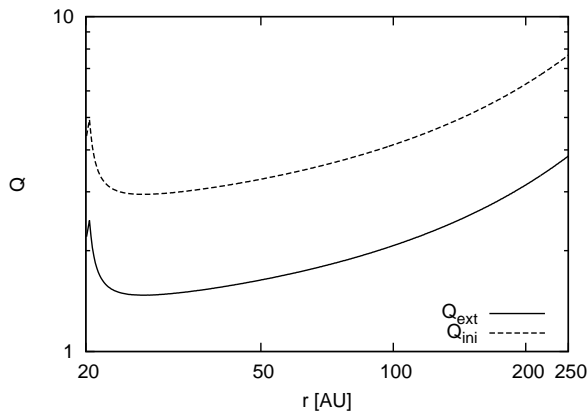}
 \caption{Distribution of $Q_{\rm ini}$ and $Q_{\rm ext}$ for model S199k005.}
 \label{fig:QiniQext} 
\end{center}
\end{figure}

We define $Q_{\rm ini}$ and $Q_{\rm ext}$ for describing the initial and typical $Q$ parameter for each model as,
\begin{equation}
 Q_{\rm ini} = \frac{\cs(T_{\rm ini})\Omega_{\rm ep,ini}}{\pi G \Sigma_{\rm ini}},
\end{equation}
\begin{equation}
 Q_{\rm ext} = \frac{\cs(T_{\rm ext})\Omega_{\rm ep,ini}}{\pi G \Sigma_{\rm ini}}.
\end{equation}
Fig. \ref{fig:QiniQext} shows the distribution of $Q_{\rm ext}$ and $Q_{\rm ini}$ for model S199k005.
Since $Q_{\rm ini} \gtrsim 2$, the protoplanetary disks are gravitationally stable at $t=0$.
In initial stage of the simulation, the temperature of the disks decreases to $\sim T_{\rm ext}$ due to the radiative cooling, and the distribution of $Q$ parameter is roughly converged to $Q_{\rm ext}$.
Then disks become gravitationally unstable for a global mode and the spiral arms are formed.

We add small level (=0.1 per cent) of white noise to the initial surface density.

\begin{table*}
\begin{center}
\caption{Model parameters and calculation results}
\label{tab:models}
\begin{tabular}{cccccccc} \hline\hline
 Models & $R_{\rm in}$ & $\Sigma_{100}$ &
 $T_{100}$ & $\kappa_{10}$ & softening length & disk-star-mass ratio &fragmentation \\ 
 & (AU)& (g cm$^{-2}$) & (K) & (cm$^{2}$ g$^{-1}$) & r &\\ \hline
S199k005 & 20 & 19.9 & 20.8 & 0.05 & 0.012 & 0.57 & No \\
S265k005 & 20 & 26.5 & 20.8 & 0.05 & 0.012 & 0.76 &Yes \\
S431Tx4 & 20 & 43.1 & 83.4 & 0.05 & 0.012 & 1.2 & No \\
S497Tx4 & 20 & 49.7 & 83.4 & 0.05 & 0.012 & 1.4 & No \\
S563Tx4 & 20 & 56.3 & 83.4 & 0.05 & 0.012 & 1.6 & No \\ 
Rin50 & 50 & 26.5 & 20.8 & 0.05 & 0.012 & 0.54 & No \\
Rin5 & 5 & 26.5 & 20.8 & 0.05 &  0.012 & 1.03 & Yes\\
S265sft003 & 20 & 26.5 & 20.8 & 0.05 & 0.03 & 0.76 &No \\ 
S298sft003 & 20 & 29.8 & 20.8 & 0.05 & 0.03 & 0.76 &Yes \\ 
\hline
\end{tabular}
\end{center}
\end{table*}

\section{Results}
\label{result}

We calculate the evolution of the disks until about $1.4\times 10^5 $ yr, which corresponds to the 10 orbital periods at $r = 100 $ au.

\subsection{No fragmentation case}
\subsubsection{Time evolution}

\begin{figure*}
 \includegraphics[width=80mm]{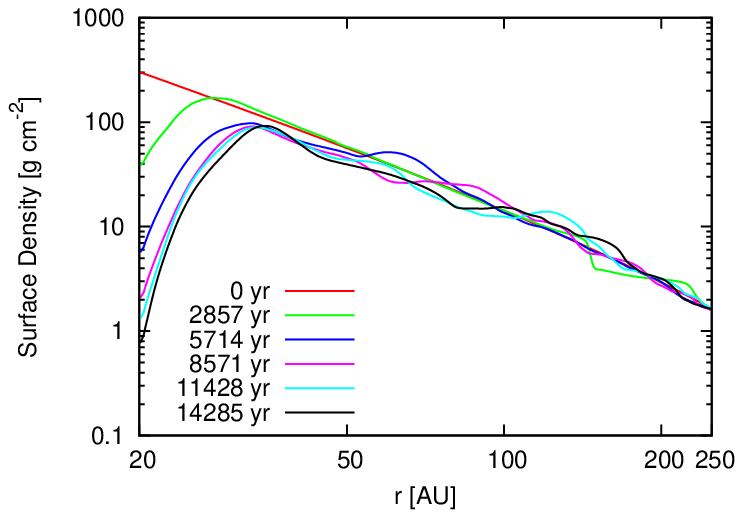}
 \includegraphics[width=80mm]{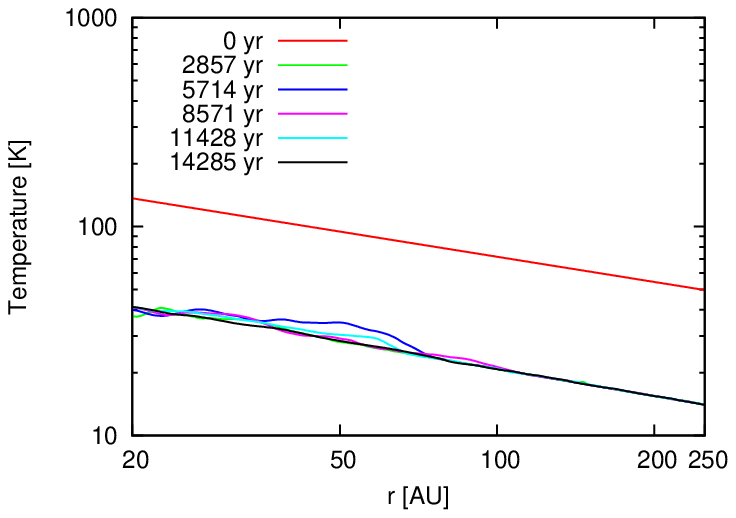}
 \includegraphics[width=80mm]{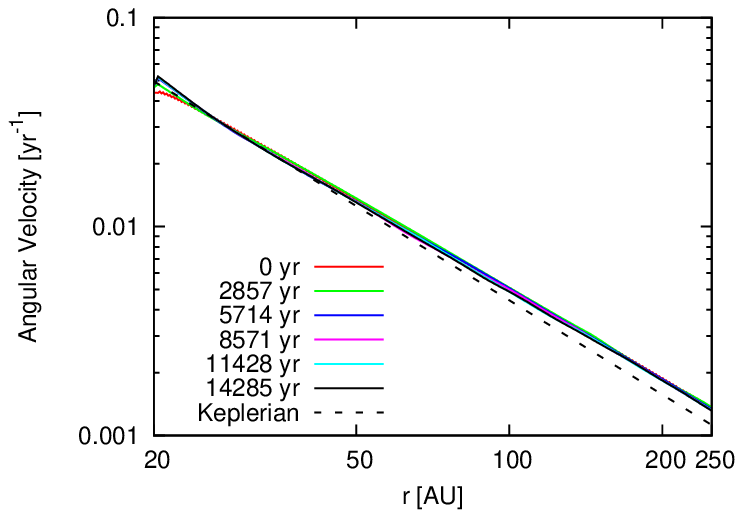}
 \includegraphics[width=80mm]{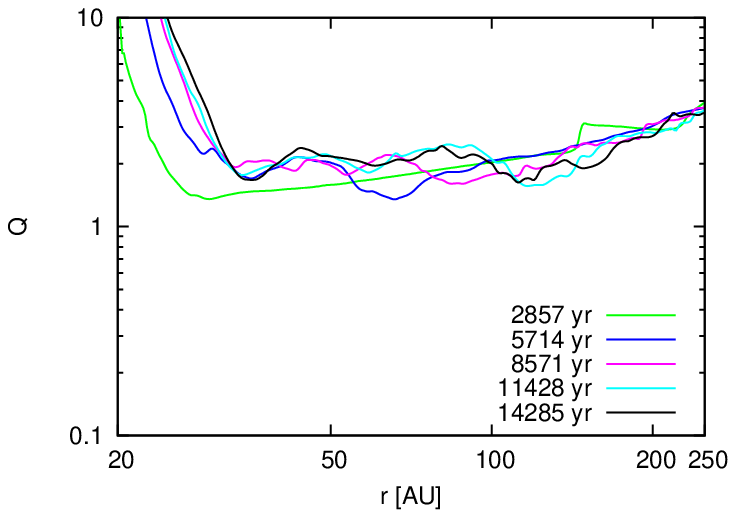}
\caption{Radial distribution of the azimuthally averaged value for $t=$ 0, 2857, 5714, 8571, 11428, 14285 yr for model S199k005.
 The top left panel is the averaged surface density, the top right panel is the averaged temperature, the bottom left panel is averaged angular frequency, and the bottom right panel is averaged Toomre $Q$ parameter.}
\label{fig:rad_ave_long_disk_2}
\end{figure*}

Fig. \ref{fig:rad_ave_long_disk_2} shows the radial distribution of the azimuthally averaged surface density, temperature, angular frequency, and Toomre $Q$ parameter for model S199k005 in Table \ref{tab:models}. 
In model S199k005, fragmentation does not happen.
Thus smooth, quasi-steady structure is realized.
The surface density in the inner region ($ r\lesssim 40$ au) decreases rapidly because of the inner boundary. 
The effect of the inner boundary is discussed in Section \ref{inner_radius}.
The averaged surface densities in the region $r \gtrsim 40$ au changes slightly from the initial surface density because of formation of spiral arms in the disk.
The temperature of the disk decreases rapidly, and the temperature distributions are roughly given by $T_{\rm ext}$ (Equation (\ref{eq:T_ext})).
The temperature around 50 au is a little bit larger than $T_{\rm ext}$ due to shock heating by spiral arms.
The averaged angular velocity is slightly different from that of Kepler rotation because of self-gravity and the pressure gradient of the disk.
The angular velocity profile does not significantly change in time.
The averaged Toomre $Q$ parameter is roughly given by $Q_{\rm ext}$ at $t= 2857$ yr and after that $Q$ is sustained for $\sim 2$. 
The tendency that the averaged $Q$ parameter remains nearly constant in self-gravitating disks is realized in many previous works.

\subsubsection{The structures of the spiral arm}
\begin{figure*}
\includegraphics[width=80mm]{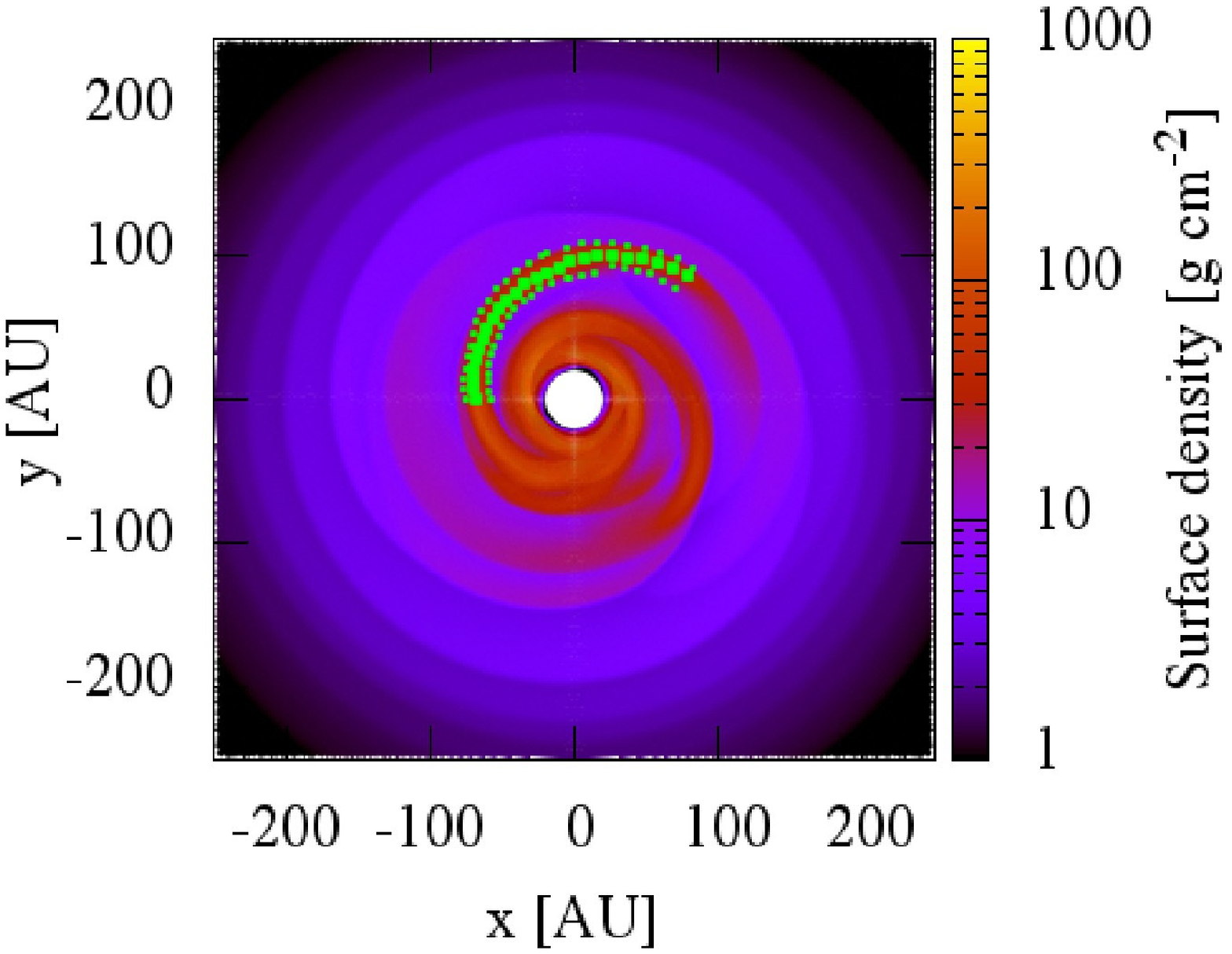}
\includegraphics[width=80mm]{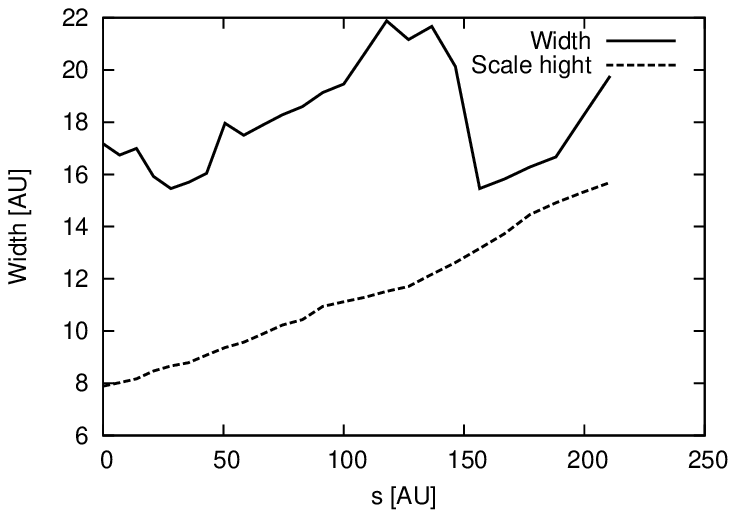}
\includegraphics[width=80mm]{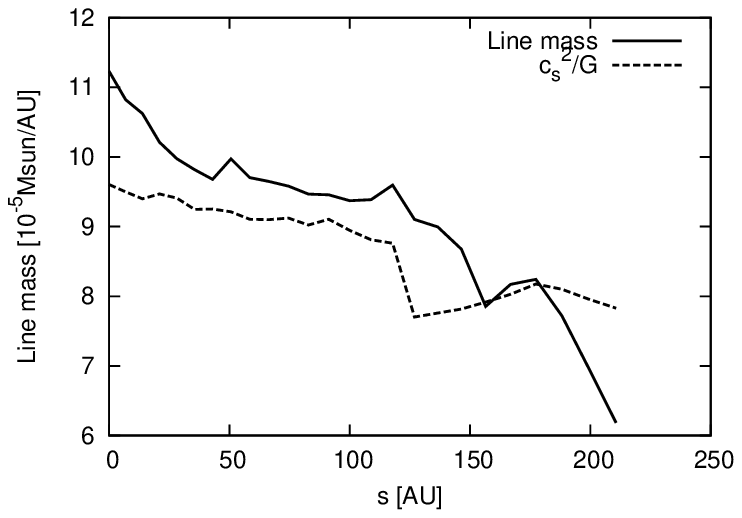}
\includegraphics[width=80mm]{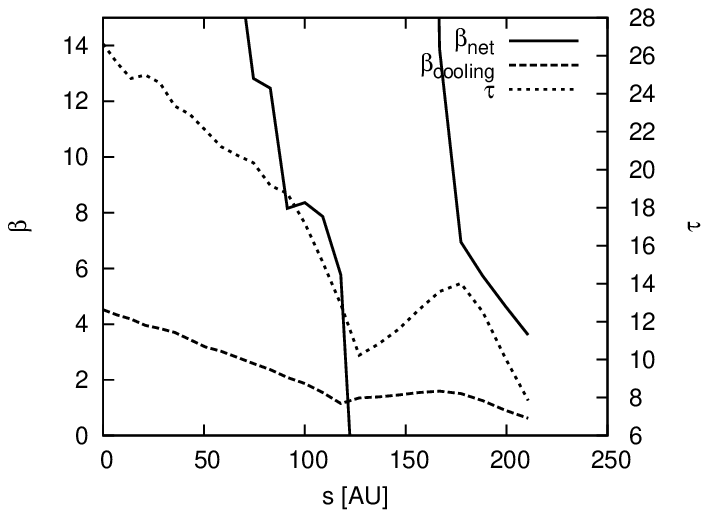}
\caption{Structure of the spiral arm at $t$=7571 years for model S199k005.
 The left top panel shows the spiral arm that we focus on by green squares.
 The right top panel shows  the width of the spiral arm and the scale height $\cs/\Omega$ evaluated at the center of the spiral.
 The left bottom panel shows the line mass of the spiral arm.
 The right bottom panel shows the normalized cooling time $\beta_{\rm cooling}$ and $\beta_{\rm net}$ and the optical depth $\tau$.}
\label{fig:spiral_not_fragment}
\end{figure*}

We perform the detailed analysis of spiral arms to investigate the condition for fragmentation.
Fig. \ref{fig:spiral_not_fragment} shows the structure of the spiral arm at $t=7571$ yr for model S199k005. 
The left top panel of Fig. \ref{fig:spiral_not_fragment} shows the map of surface density.  The green squares trace the spiral arm that we focus on.
The large green squares show the centers of the spiral arm and the small green squares show the edge of the spiral arm.
The centers of the spiral arm are obtained by following the peak of the surface density. 
The separation between the large green points is $0.1r$. The edge of the spiral is defined as the point where the surface density is 0.3 times the surface density  at the center of spiral arm.
The other panels in Fig. \ref{fig:spiral_not_fragment} show the distribution of the physical values along the spiral arm.
The horizontal axis $s$ is the distance along the spiral arm.
The origin of $s$ is the large green point whose radius is the minimum.
The right top panel shows the width of the spiral arm and the scale height $\cs/\Omega$ evaluated at the center of the spiral.
The width of the spiral arm is about twice the scale height for $s \lesssim 150$ au and comparable to the scale height for $s \gtrsim 150$ au.
The left bottom panel shows the line mass of the spiral arm.
The line mass of the spiral arm is approximately given by $\cs^2 / G$.
Since the line mass is smaller than the critical line mass $2\cs^2/G$, the spiral is supported by the pressure against the self-gravity in the direction perpendicular to the spiral arm.
The right bottom panel shows the normalized cooling time and the optical depth $\tau$.
We show two different normalized cooling times that are defined as follows;
\begin{equation}
 \beta_{\rm net} \equiv
 \frac{E\Omega}{\Lambda_{\rm c}},
\end{equation}
\begin{equation}
 \beta_{\rm cooling} \equiv
  E\Omega\frac{3\left(\frac{1}{4}\tau^2+\frac{1}{\sqrt{3}}\tau+\frac{2}{3}\right)}
{8\sigma T^4 \tau},\label{beta_cooling}
\end{equation}
where $\beta_{\rm cooling}$ indicates the
time scale by the radiative cooling, and in $\beta_{\rm net}$ include the effect of external heating.
Since the cooling is efficient and the temperature of the spiral arm $T$ is very close to or smaller than $T_{\rm ext}$, $\beta_{\rm net}$ is very large or negative at $s \sim 150$ au.
The spiral arm is optically thick and the cooling time is roughly proportional to $T^{-1}$.

\subsubsection{Comparison with cooling criterion}
According to the cooling criterion, the spiral arm fragments when the cooling time is small enough.
Although there is a controversy on the critical value of the normalized cooling time $\beta_{\rm crit}$, the proposed values are all larger than 3. 
In the spiral arm shown in Fig. \ref{fig:spiral_not_fragment}, however, satisfies $\beta_{\rm cooling} < 3$ for $s \gtrsim 50$ au but does not fragment.
This result clearly contradicts the cooling criterion proposed in previous works.
One may argue that $\beta _{\rm net }$ should be used for the criterion.
In this case, this disk does not satisfy the criterion. 
However, we will show examples that satisfy $\beta_{\rm net }$ and does not fragment in Section \ref{counter-examples}.
In this way, we can see that the cooling criterion is not always valid.

\subsection{Fragmentation case}
To investigate the fragmentation process of the disk, we perform the numerical simulation in which the initial disk mass is larger than that of model S199k005.
\subsubsection{Time evolution}
\begin{figure*}
 \includegraphics[width=80mm]{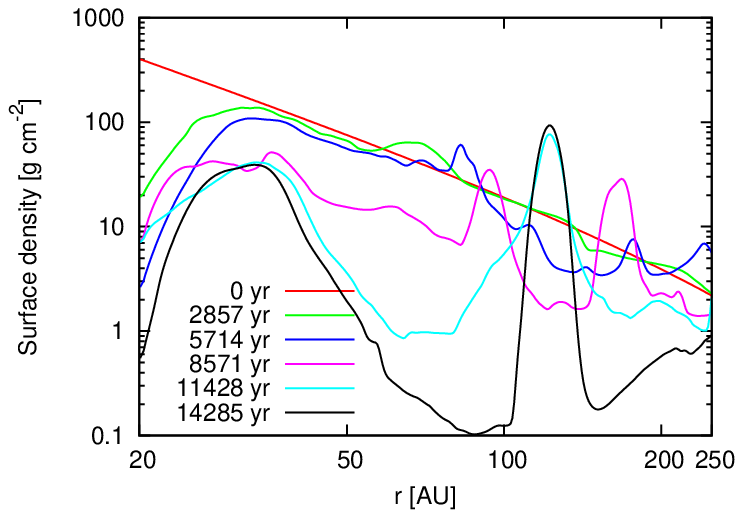}
 \includegraphics[width=80mm]{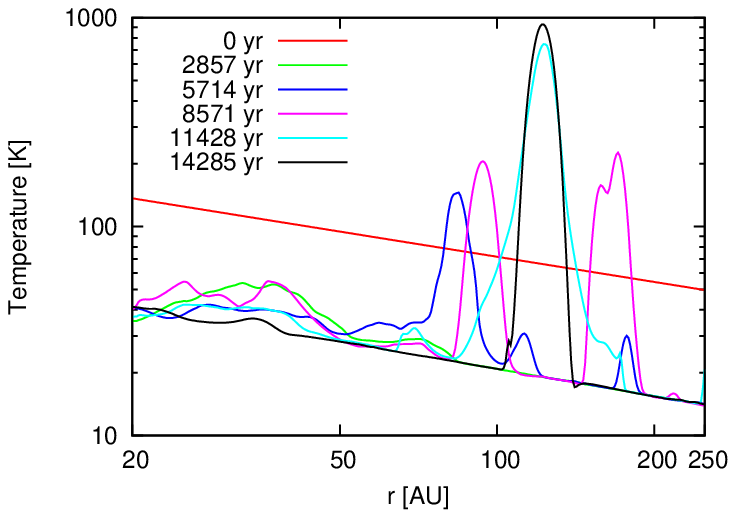}
 \includegraphics[width=80mm]{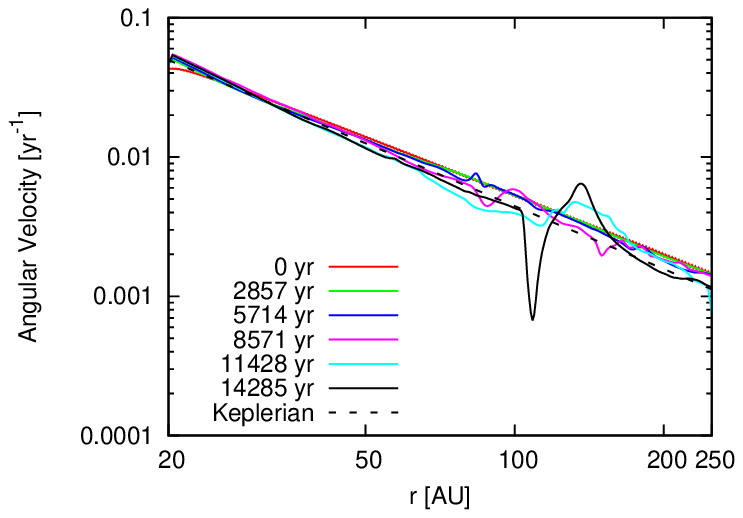}
 \includegraphics[width=80mm]{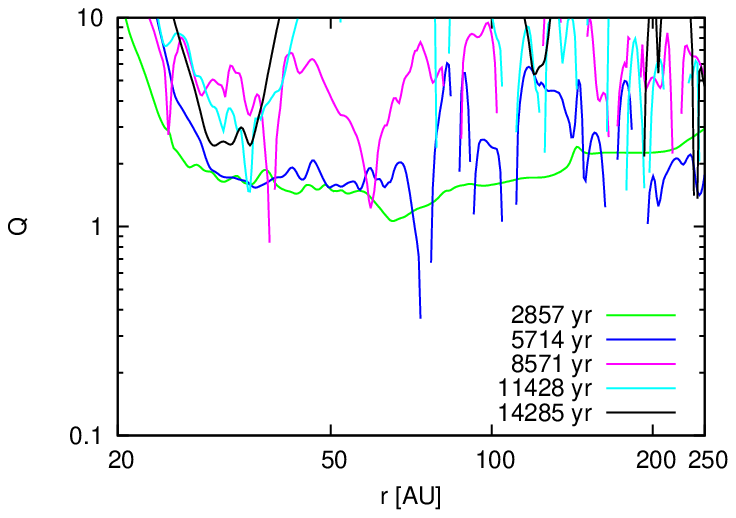}
\caption{Radial distribution of the azimuthally averaged value for $t=$ 0, 2857, 5714, 8571, 11428, 14285 yr for model S265k005, whose initial surface density is larger than that of model S199k005.
 The top left panel is the averaged surface density, the top right panel is the averaged temperature, the bottom left panel is the  averaged angular frequency, and  the bottom right panel is the averaged Toomre $Q$ parameter.}
\label{fig:rad_ave_long_disk}
\end{figure*}

In model S265k005, whose initial disk mass is about 1.3 times larger than that of S199k005, the fragmentation occurs in the spiral arms.
Fig. \ref{fig:rad_ave_long_disk} shows the radial distribution of the azimuthally averaged surface density, temperature, angular velocity and Toomre $Q$ parameter. 
The fragments are formed between $t=2857$ and 5714 yr. 
The surface density in the inner region ($ r\lesssim 40$ au)  decreases by the rapid accretion through the inner boundary. 
However, since the fragmentation occurs at the radius $r \sim 100$ au, the effect of the inner boundary does not affect the fragmentation. 
After the fragmentation occurs, the angular momentum is efficiently transferred by the fragments, promoting rapid gas accretion onto the central star.
After the fragmentation, the temperatures of the fragments are much higher than $T_{\rm ext}$, but the temperature of the disk is almost same as $T_{\rm ext}$.
The opacity model (Equation (\ref{eq:opacity})) is relevant when $T < 200$ K, and the disk temperature satisfy this condition except in the fragments. 
The averaged distribution of $Q$ parameter is roughly given by $Q_{\rm ext}$ before the fragmentation.
On the other hand, after the fragmentation, the gas of the disk is disturbed by the fragments and the square of epicycle frequency becomes negative in some regions. Thus $Q$ is not defined in such regions.

\begin{figure}
 \includegraphics[width=80mm]{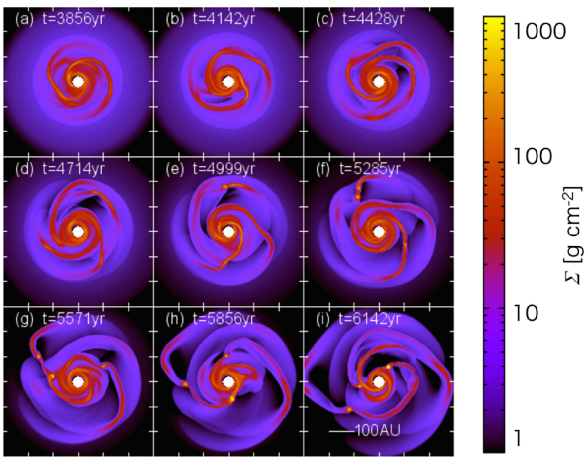}
 \caption{Time sequence of the surface density for model S265k005.}
\label{fig:9maps_Sigma_fragment}
\end{figure}

\begin{figure}
 \includegraphics[width=80mm]{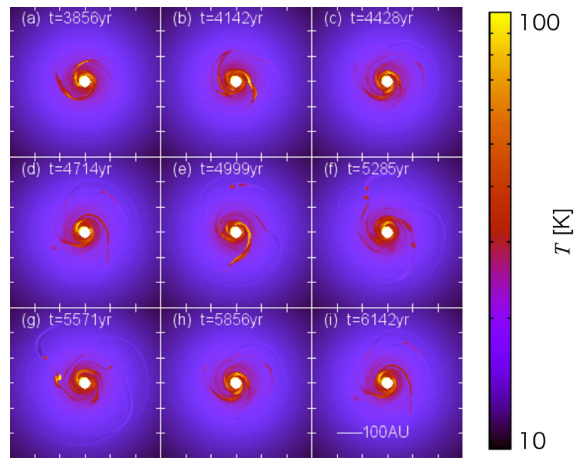}
 \caption{Time sequence of the temperature distribution for model S265k005.}
\label{fig:9maps_T_fragment}
\end{figure}

Fig. \ref{fig:9maps_Sigma_fragment} shows the time sequence of the surface density of the disk for model S265k005.
The circumference of the white filled circle shows the inner boundary.
In Fig. \ref{fig:9maps_Sigma_fragment}, the three fragments are formed in the two spiral arms.  
Fig. \ref{fig:9maps_T_fragment} shows the time sequence of the temperature distribution in the disk for model S265k005 at the same epochs as in Fig. \ref{fig:9maps_Sigma_fragment}.
The structures of the spiral arms are less clear in the temperature plot than that in the surface density plot because the spiral heating is weak.
In particular, the structures of the temperature are almost axisymmetric in the region $r\gtrsim 100$ AU indicating that the temperature is determined by external heating. 

\subsection{Condition for fragmentation of the spiral arms}
\begin{figure}
\includegraphics[width=80mm]{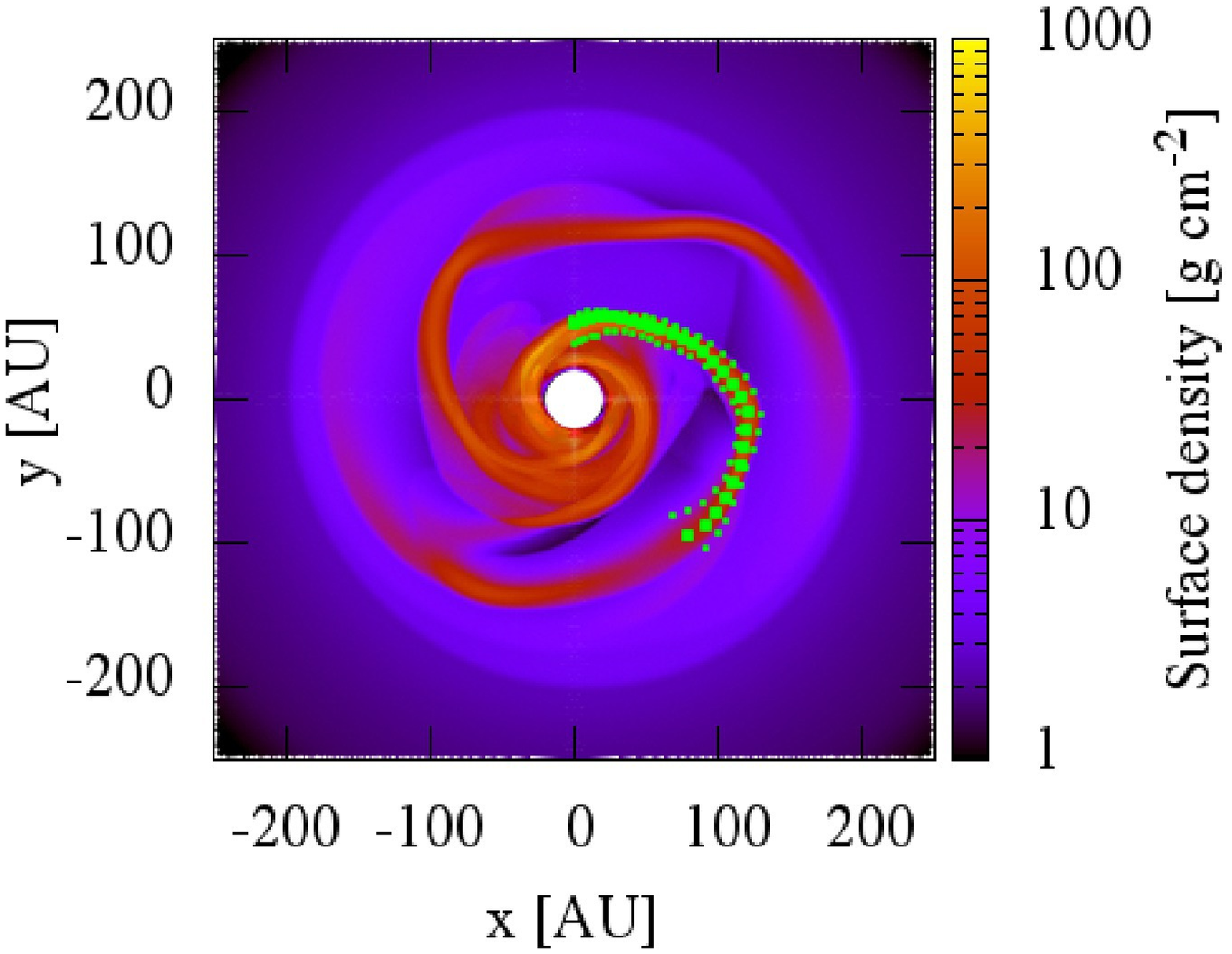}
\includegraphics[width=80mm]{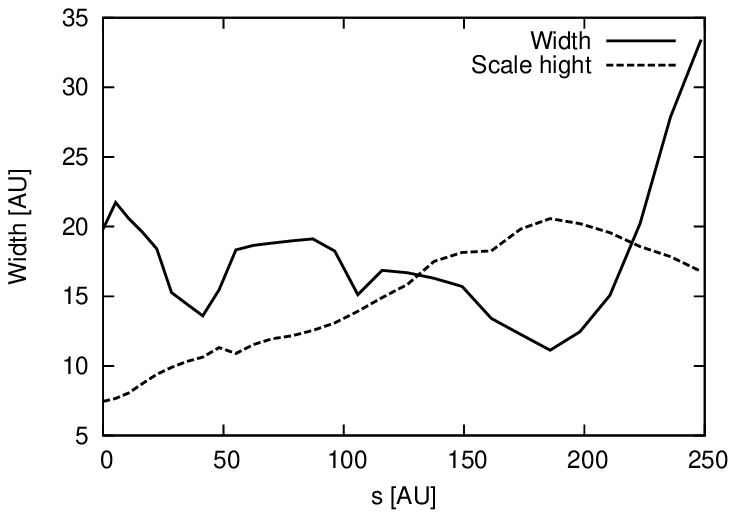}
\includegraphics[width=80mm]{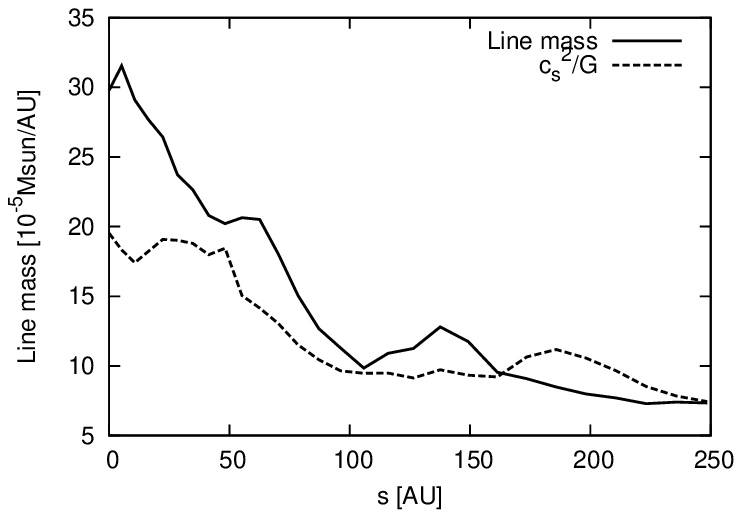}
\includegraphics[width=80mm]{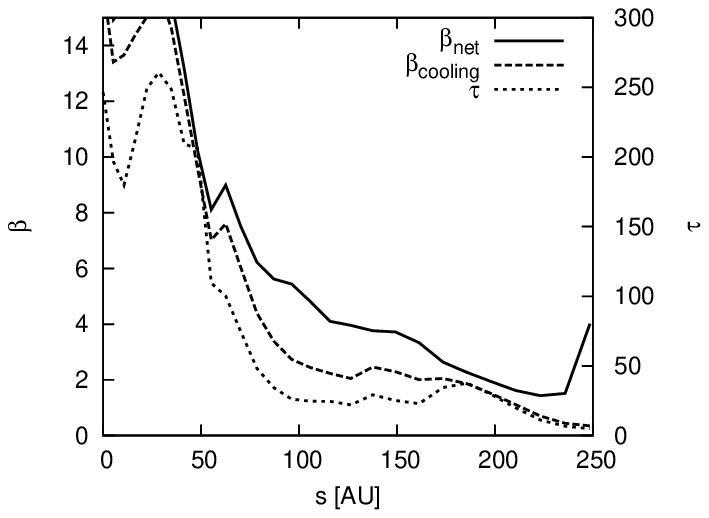}
\caption{Structure of the spiral arm at $t$=4285 years for model S265k005.
}
\label{fig:spiral_fragment}
\end{figure}

Fig. \ref{fig:spiral_fragment} shows the structure of the spiral arm at 4285 yr for model S265k005. 
This spiral arm fragments at $\sim 5\times 10^3$ yr (Fig. \ref{fig:9maps_Sigma_fragment}).
The width of the spiral arm is comparable to the scale height and the line mass of the spiral arm is approximately given by $\cs^2 /G$.
In the region $s \gtrsim 100$ au, the normalized cooling time in the spiral arm is $\beta \sim 2$ and similar to that of model S199k005 shown in Fig \ref{fig:spiral_not_fragment}.
Thus, it is difficult to distinguish these two spiral arms by the normalized cooling time. 

\begin{figure}
\includegraphics[width=80mm]{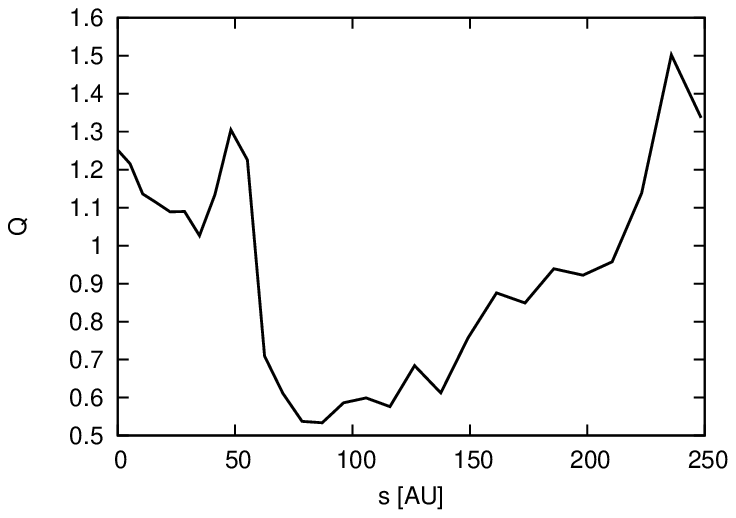}
\includegraphics[width=80mm]{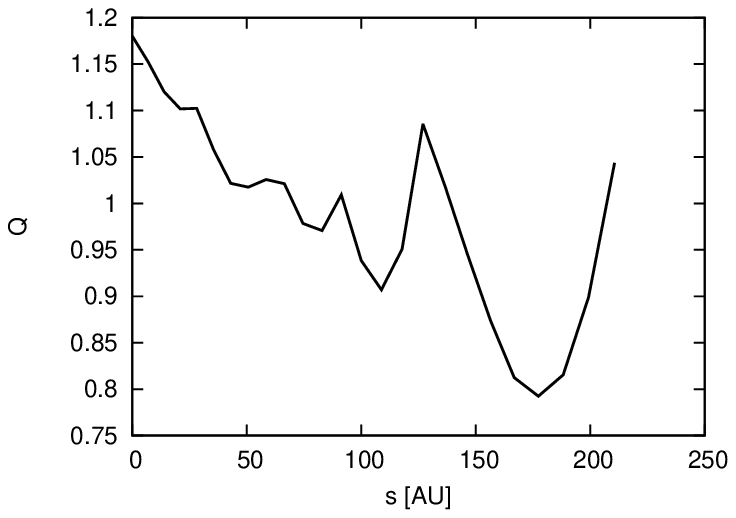}
\caption{Toomre $Q$ parameters in the spiral arms. The top panel shows $Q$ in the spiral arm shown in Fig. \ref{fig:spiral_fragment} and the bottom panel shows $Q$ in the spiral arm shown in Fig.\ref{fig:spiral_not_fragment}, respectively.
The spiral arm shown in Fig. \ref{fig:spiral_fragment} does not fragment and the spiral arm shown in Fig.\ref{fig:spiral_not_fragment} fragment.
These results are consistent with the criterion for the fragmentation: $Q <0.6$.
 }
\label{fig:arms_s-Q}
 \end{figure}

What is an essential parameter of fragmentation?
We propose that a local $Q$ parameter in a spiral arm is the critical parameter that controls fragmentation of spiral arms.
Fig. \ref{fig:arms_s-Q} shows $Q$ parameter in the spiral arms shown in Fig. \ref{fig:spiral_fragment} and \ref{fig:spiral_not_fragment}.
The minimum $Q$ parameter in the spiral arm for model S265k005 is about 0.5 and the minimum $Q$ parameter in the spiral arm for model S199k005 is about 0.8.
From the results of numerical simulations that we perform in this work, we find that the spiral arms fragment when $Q<0.6$ is satisfied.
We have examined this condition by using many runs that include both fragmenting spiral arms and non-fragmenting spiral arms and found that this condition $Q\lesssim 0.6$ for the fragmentation of the spiral arms is valid for all the spiral arms that we examined. 
These results indicate that the $Q$ parameter {\it in the spiral arms} is an essential parameter for the fragmentation process of disks.
In Section \ref{linear_ana}, we show that the condition for $Q$ parameter can be derived from the linear analysis for the gravitational instability of the tightly wound spiral arm.
\subsection{Relation between normalized cooling time and fragmentation}
\label{largeT}
In this section, we investigate the relation between the normalized cooling time and fragmentation and show that normalized cooling time does not determine the condition of fragmentation. 
We perform the numerical simulation of hotter disks than that of model S199k005 and S256k005 to investigate whether the gravitational fragmentation occurs  when $\beta$ in the spiral arms decrease.
Note that $\beta_{\rm cooling } \propto T^{-1}$ in optically thick spiral arms (see Equation (\ref{beta_cooling})).
The temperatures of the disks are roughly equal to $T_{\rm ext}$.
Thus we can control the temperature of the spiral arms by $T_{\rm ext}$.
In model S431Tx4, S497Tx4, and S563Tx4, the external temperature is higher by a factor of 4 than that of model S265k005. 
The surface densities are also elevated  to make the disks gravitationally unstable.

\begin{figure}
\includegraphics[width=80mm]{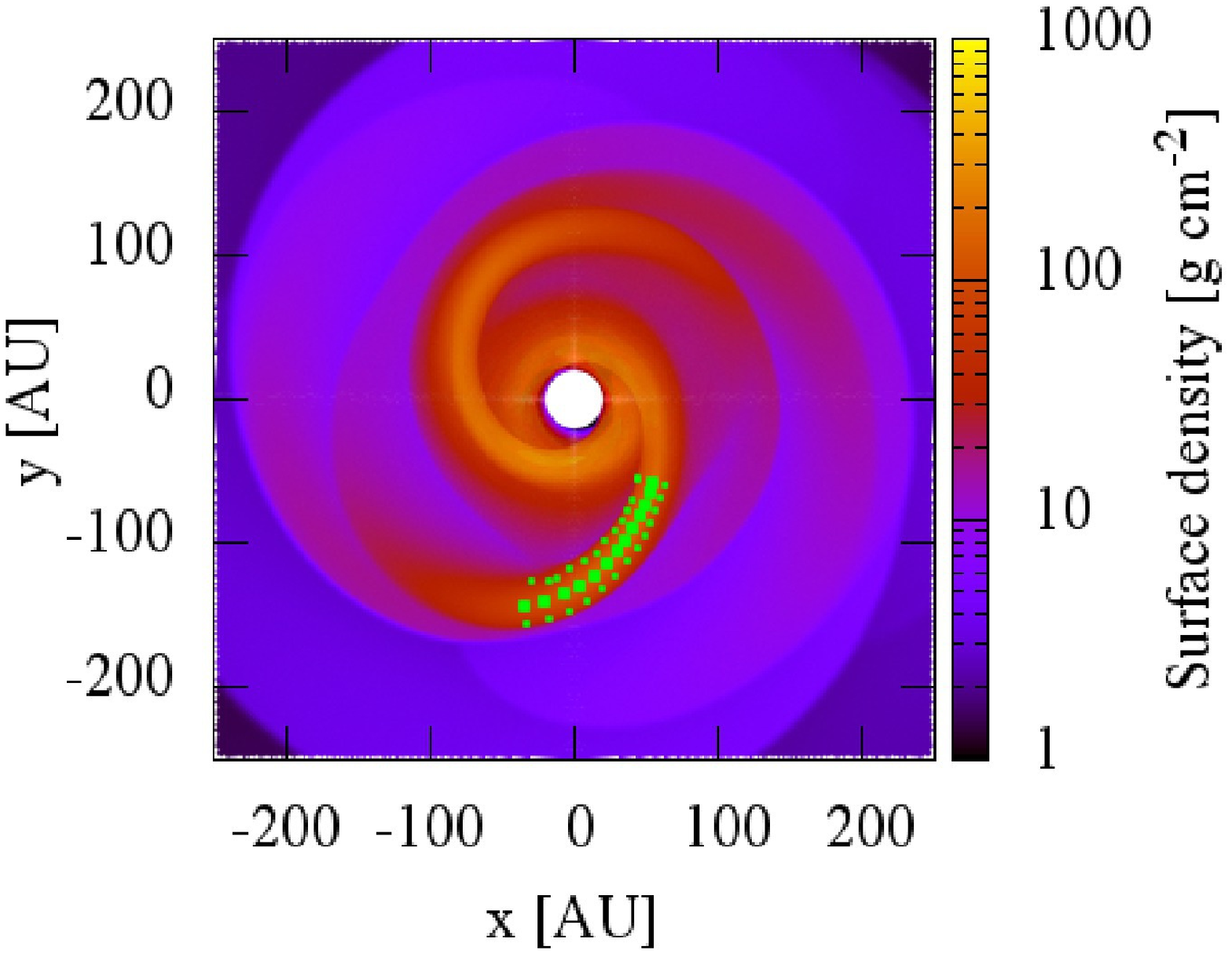}
\includegraphics[width=80mm]{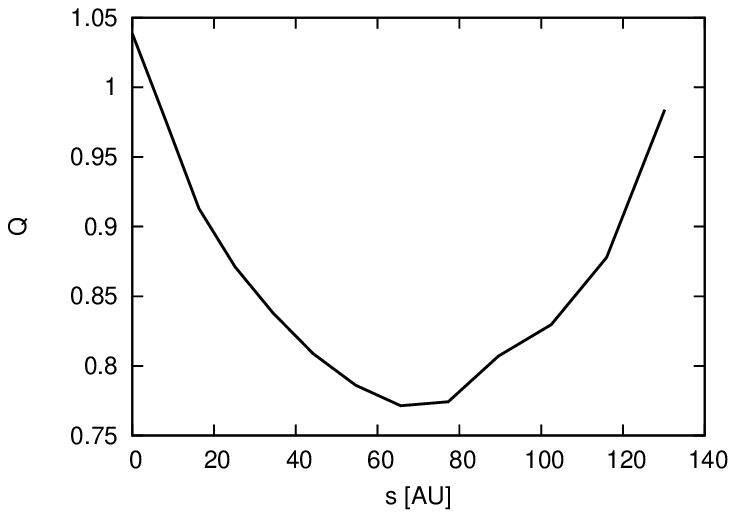}
\includegraphics[width=80mm]{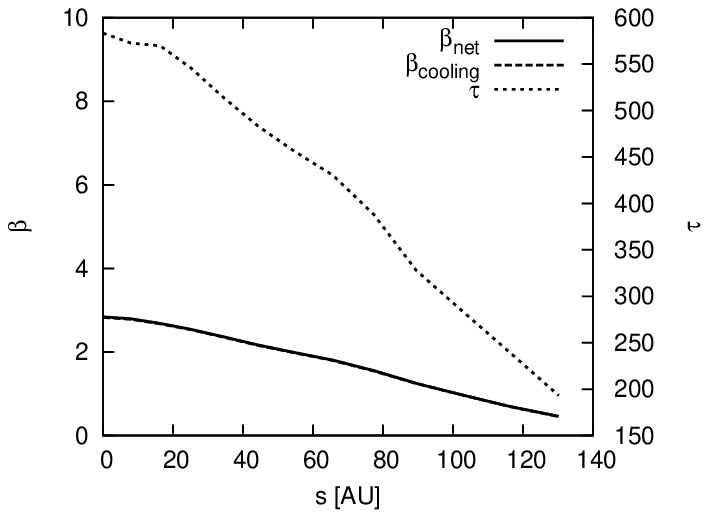}
\caption{Results of model S563Tx4.  $t$=4999 years. 
 This spiral arm  does not fragment.
  Thus, this result is also consistent with the criterion for the fragmentation: $Q <0.6$.}
\label{fig:Tx4sigma0030}
\end{figure}

Fig. \ref{fig:Tx4sigma0030} shows the spiral structure of model S563Tx4 at $t$=4999 years. This spiral arm does not fragment even with small $\beta$.
The width of the spiral arm is larger than 20 AU.
The $Q$ parameter at the center of the spiral is larger than 0.75 and the absence of fragmentation is consistent with our criterion for the fragmentation $Q \lesssim 0.6$.
Since the temperature of the spiral is large, the cooling rate is also large. 
As  a  result, the minimum normalized cooling time in the spiral arm is about unity.
Although $\beta_{\rm cooling}$ and $\beta _{\rm net}$ in the spiral arm in model S563Tx4 is smaller than that of model S265k005, the spiral arms in model S563Tx4 do not fragment and the spiral arms in model S265k005 fragment.Note that $Q_{\rm ext}$ of the model S563Tx4 is smaller than that of model S265k005 since the initial surface density of the disk in model S563Tx4 is about 2.1 times larger than that of model S265k005.

\cite{2011MNRAS.418.1356R} gives the relation between $\beta_{\rm crit}$ and external irradiation by the local two-dimensional numerical simulations.
Their results suggest that $\beta_{\rm crit}$ increases as $Q_{\rm ext}$ decreases.
On the other hand, the results of S265k005 and S563Tx4 suggest that the $\beta_{\rm crit}$ decreases as $Q_{\rm ext}$ decreases.
Thus,  our results conflict with the results of local simulations done by \cite{2011MNRAS.418.1356R}.
This indicates that the critical value of normalized cooling time depends not only on $Q_{\rm ext}$ but also on other parameters of the disk, for example, the temperature, the total disk mass, and the radius.
Therefore, it seems quite difficult to obtain the critical cooling time that can be universally applied for all disks.

\subsection{Effect of inner boundary}
\label{inner_radius}
\begin{figure}
\includegraphics[width=80mm]{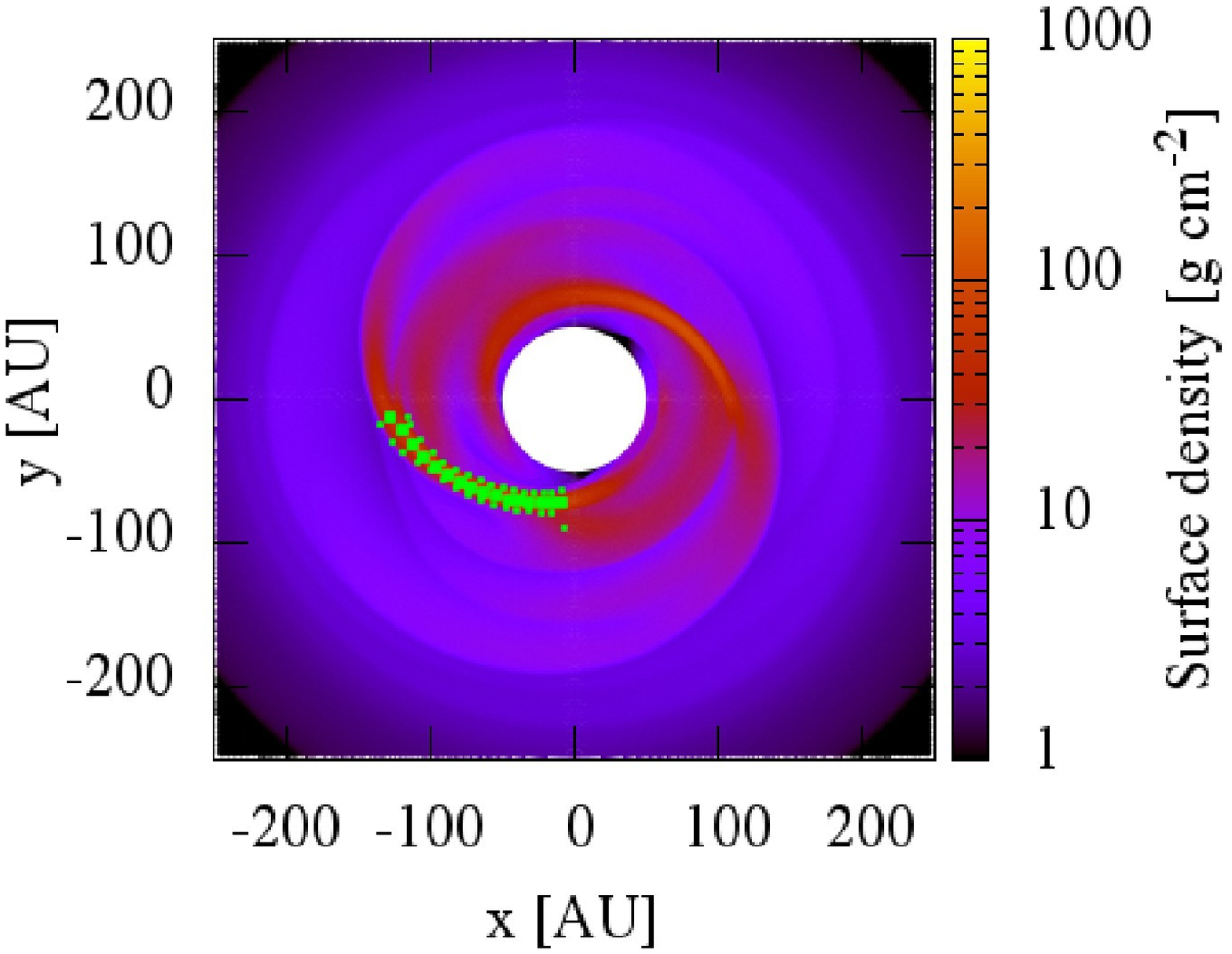}
\includegraphics[width=80mm]{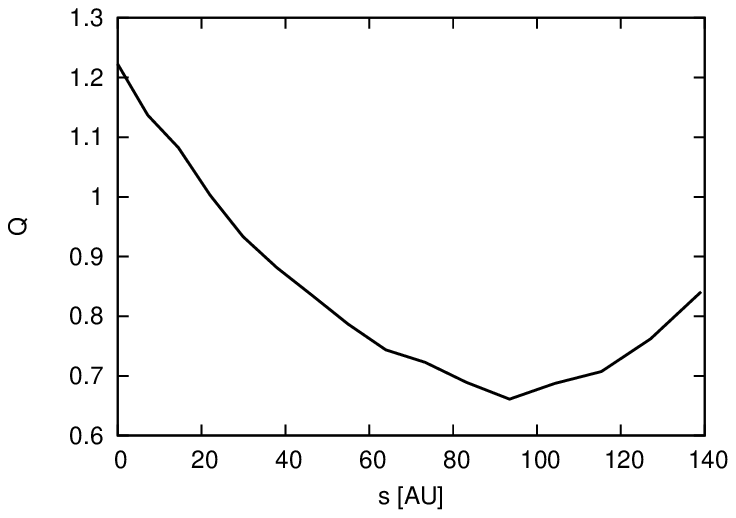}
\caption{Structure of the spiral arm at $t$=7142 years for model Rin50.
 This spiral arm  does not fragment.
  Thus, this result is also consistent with the criterion for the fragmentation: $Q <0.6$.}
\label{fig:spiral_Rin50}
\end{figure}

 In the simulations performed in this work, the surface density in the inner region decreases quickly in the early evolution because of the rapid accretion through the inner boundary. 
  To investigate the effect of our inner boundary, we performed simulations in which the inner radii are different from model S265k005.  
  Those runs are labelled Rin5 and Rin50.
The inner radii of models Rin5 and Rin50 are 5 au and 50 au respectively, while the inner radius of model S265k005 is 20 au.
The fragmentation does not occur if the inner radius is very large (model Rin50). 
Fig. \ref{fig:spiral_Rin50} shows the structures of the spiral arm for model Rin50. 
In this case, the minimum $Q$ in the spiral arm is $\sim 0.7$ and the spiral arms in model Rin50 do not satisfy the condition of the fragmentation $Q\lesssim 0.6$. 
  In this way, the result of simulation depends on the inner boundary condition but our condition for the fragmentation still remains valid.
Thus we see that our condition for the fragmentation is not affected by the inner radius and the inner boundary condition.

\subsection{Influence of disk thickness}
\label{softening_length}
\begin{figure}
\includegraphics[width=80mm]{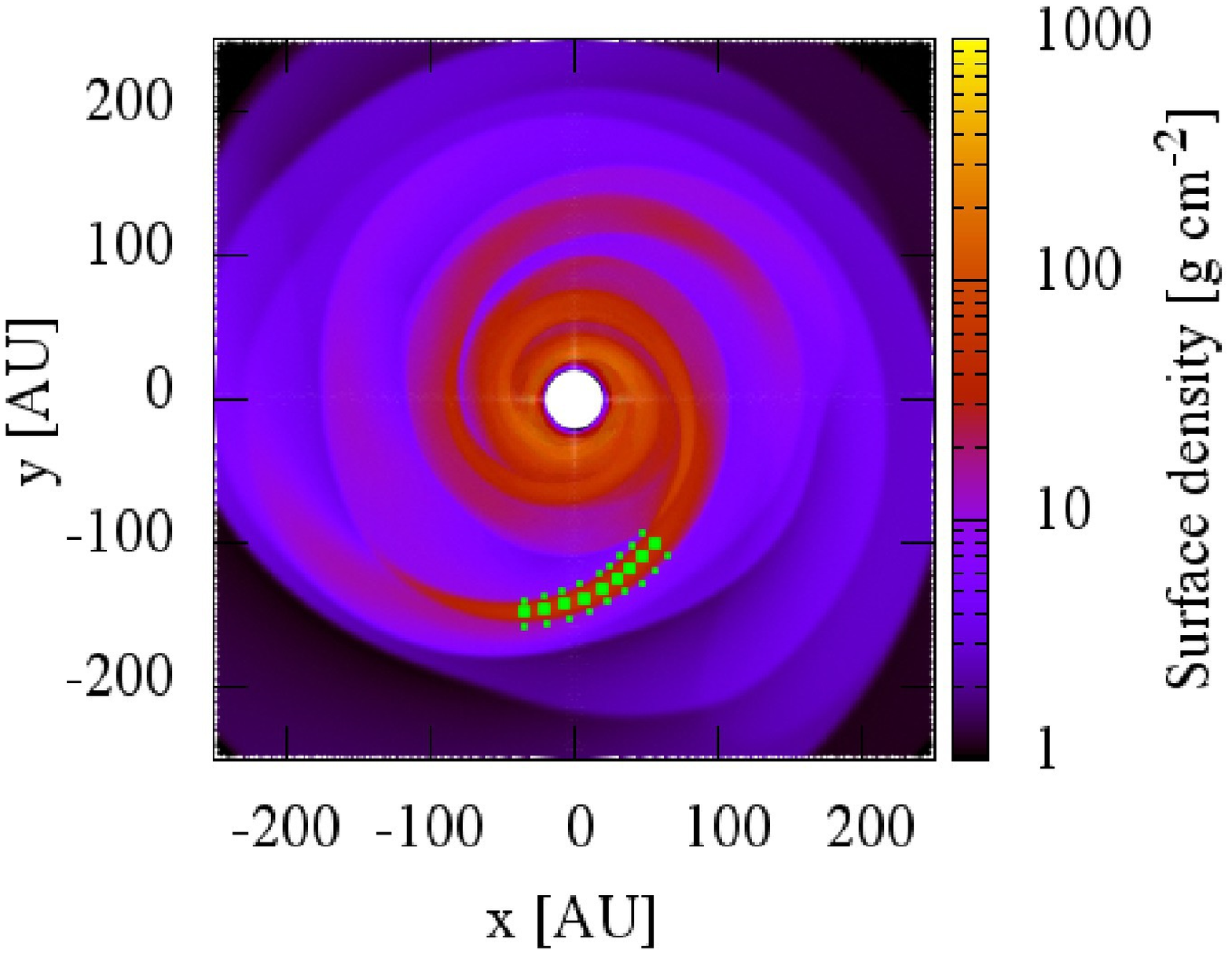}
\includegraphics[width=80mm]{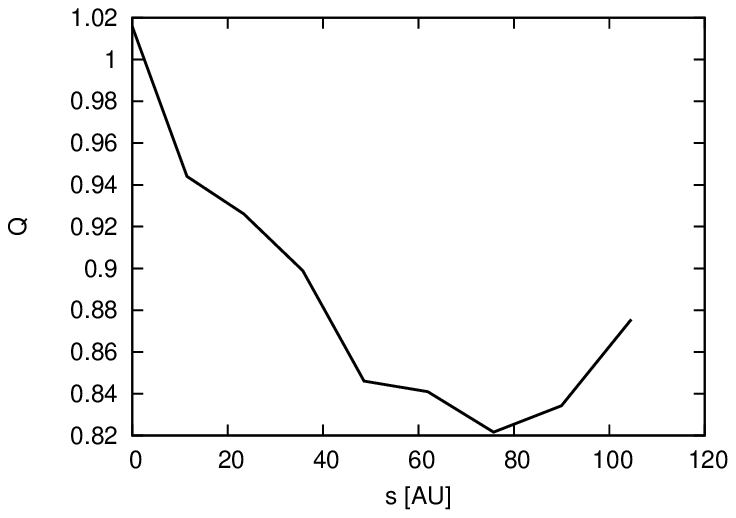}
\caption{Structure of the spiral arm at $t$=10428 years for model S265sft003.
 This spiral arm  does not fragment.
Thus, this result is also consistent with the criterion for the fragmentation: $Q <0.6$.}
\label{fig:spiral_SMOOTH05_not_frag}
\end{figure}

\begin{figure}
\includegraphics[width=80mm]{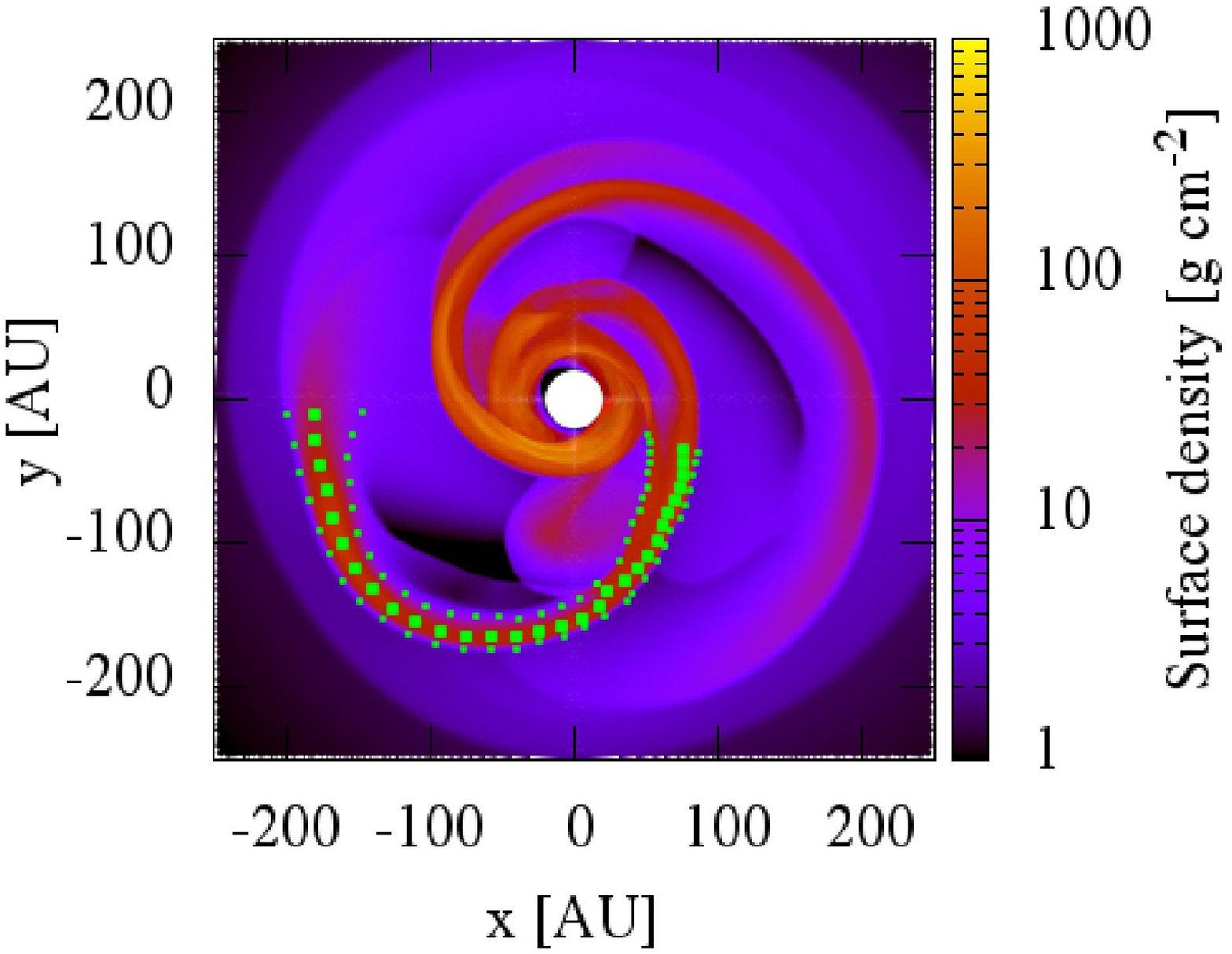}
\includegraphics[width=80mm]{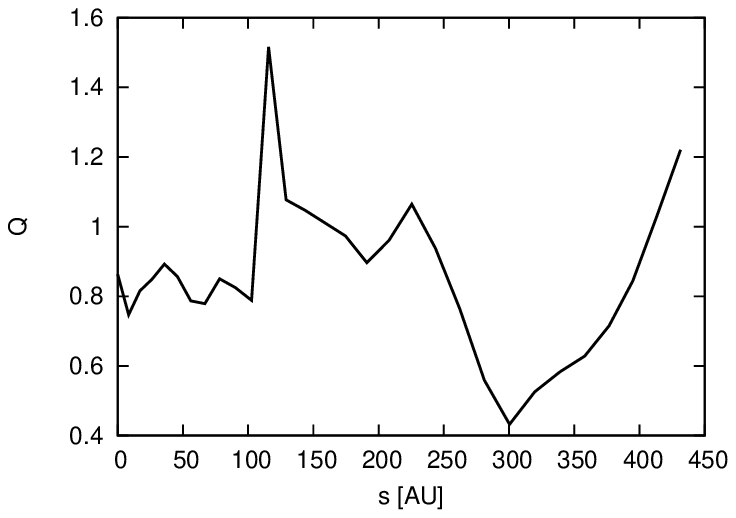}
\caption{Structure of the spiral arm at $t$=5285 years for model S298sft003. This spiral arm fragments.
  Thus, this result is also consistent with the criterion for the fragmentation: $Q <0.6$.}
\label{fig:spiral_SMOOTH05_frag}
\end{figure}

Since we use the two-dimensional numerical simulation code, we do not calculate the vertical structure of the disk.
In general, the thickness of the disk affects the self-gravity of the disk.
We model the effect of the thickness of the disk by using the softening length of the self-gravity.
By including the gravitational  softening, the small scale gravitational interaction within the softening length is smoothed.
Using this property, we can mimic the smoothing effect introduced by the scale height on the gravitational force.
In the case that we adopt the softening length 0.012$r$, the critical $Q$ parameter for the fragmentation is $\sim 0.6$, as discussed so far.
To investigate the dependence on the softening length, we perform the simulations for the softening length 0.03$r$ (model S265sft003, S298sft003).
The parameters of model S265sft003 are the same as S265k005 except the softening length, but fragmentation does not occur in model S265sft003.
This result indicates that fragmentation is difficult when the softening length is large.
Fig. \ref{fig:spiral_SMOOTH05_not_frag} shows the structures of the spiral arm of model S265sft003.
Since the minimum value of $Q$ is larger than 0.6, this result is consistent with our condition for the fragmentation.
Fig. \ref{fig:spiral_SMOOTH05_frag} shows the structures of the spiral arm of model S298sft003 in which the fragmentation occurs. 
In this case, the minimum value of $Q$ is about $0.5$.
Thus this spiral arm satisfies our criterion.
Therefore, the results are also consistent with our condition for the fragmentation. 

In this work, we investigate the effect of the thickness of the disks on the conditions for the fragmentation by using the softening length.
  The pre-defined softening length, however, cannot completely capture the physics of finite thickness disks.
\cite{2015MNRAS.451.3987Y} suggests that $\beta_{\rm crit}$ may be underestimated in two-dimensional numerical simulation.
Therefore, we need the three-dimensional simulation to precisely evaluate the more realistic value of the critical $Q$ for the fragmentation of spiral arms.


\subsection{Dependence on opacity}
\label{effect_of_cooling}
\begin{table*}
\begin{center}
\caption{Model parameters and calculation results}
\label{tab:models2}
\begin{tabular}{c}
\begin{minipage}{0.5\hsize}
\begin{center}
\begin{tabular}{cccc} \hline\hline
 Model & $\Sigma_{100}$ &
  $\kappa_{10}$ & fragmentation \\ 
 & (g cm$^{-2}$) & (cm$^{2}$ g$^{-1}$) &\\ \hline
S099k00005 & 9.9  & 5$\times 10^{-4} $ &  No\\
S133k00005 & 13.3 & 5$\times 10^{-4} $ &  No\\
S166k00005 & 16.6 & 5$\times 10^{-4} $ &  No\\
S199k00005 & 19.9 & 5$\times 10^{-4} $ &  Yes\\
S232k00005 & 23.2 & 5$\times 10^{-4} $ &  Yes\\
S099k0005  & 9.9  & 5$\times 10^{-3} $ &  No\\
S133k0005  & 13.3 & 5$\times 10^{-3} $ &  No\\
S166k0005  & 16.6 & 5$\times 10^{-3} $ &  No\\
S199k0005  & 19.9 & 5$\times 10^{-3} $ &  Yes\\
S232k0005  & 23.2 & 5$\times 10^{-3} $ &  Yes\\
S265k0005  & 26.5 & 5$\times 10^{-3} $ &  Yes\\
S298k0005  & 29.8 & 5$\times 10^{-3} $ &  Yes\\
S166k001   & 16.6 & 0.01  &  No\\
S199k001   & 19.9 & 0.01  &  No\\
S232k001   & 23.2 & 0.01  &  Yes\\
S265k001   & 26.5 & 0.01  &  Yes\\
S298k001   & 29.8 & 0.01  &  Yes\\
S166k0025  & 16.6 & 0.025  &  No\\
S199k0025  & 19.9 & 0.025  &  Yes\\
S232k0025  & 23.2 & 0.025  &  Yes\\
S265k0025  & 26.5 & 0.025  &  Yes\\
S298k0025  & 29.8 & 0.025  &  Yes\\
S166k005   & 16.6 & 0.025  &  No\\
S215k005   & 21.5 & 0.05 & Yes \\
S232k005   & 23.2 & 0.05 & No \\
\hline
\end{tabular} 
\end{center} 
\end{minipage}

\begin{minipage}{0.5\hsize}
 \begin{center}
\begin{tabular}{cccc} \hline\hline
 Model & $\Sigma_{100}$ &
  $\kappa_{10}$ & fragmentation \\ 
 & (g cm$^{-2}$) & (cm$^{2}$ g$^{-1}$) &\\ \hline
S248k005   & 24.8 & 0.05 & No \\
S298k005   & 29.8 & 0.05  &  Yes\\
S331k005   & 33.1 & 0.05  &  Yes\\
S166k01    & 16.6 & 0.1  &  No\\
S199k01    & 19.9 & 0.1  &  No\\
S232k01    & 23.2 & 0.1  &  Yes\\
S265k01    & 26.5 & 0.1  &  Yes\\
S298k01    & 29.8 & 0.1  &  Yes\\
S331k01    & 33.1 & 0.1  &  Yes\\
S166k025   & 16.6 & 0.25  &  No\\
S199k025   & 19.9 & 0.25  &  No\\
S232k025   & 23.2 & 0.25  &  No\\
S265k025   & 26.5 & 0.25  &  Yes\\
S298k025   & 29.8 & 0.25  &  Yes\\
S331k025   & 33.1 & 0.25  &  Yes\\
S233k05    & 23.3 & 0.5  &  No\\
S265k05    & 26.5 & 0.5  &  No\\
S298k05    & 29.8 & 0.5  &  Yes\\
S331k05    & 33.1 & 0.5  &  Yes\\
S364k05    & 36.4 & 0.5  &  Yes\\
S298Ad     & 29.8 & Adiabatic  &  No\\
S331Ad     & 33.1 & Adiabatic  &  Yes\\
S364Ad     & 36.4 & Adiabatic  &  Yes\\
S398Ad     & 39.8 & Adiabatic  &  Yes\\
S421Ad     & 42.1 & Adiabatic  &  Yes\\
\hline
\end{tabular}
 \end{center}
\end{minipage}

\end{tabular}
\end{center}
\end{table*}

\begin{figure}
\includegraphics[width=80mm]{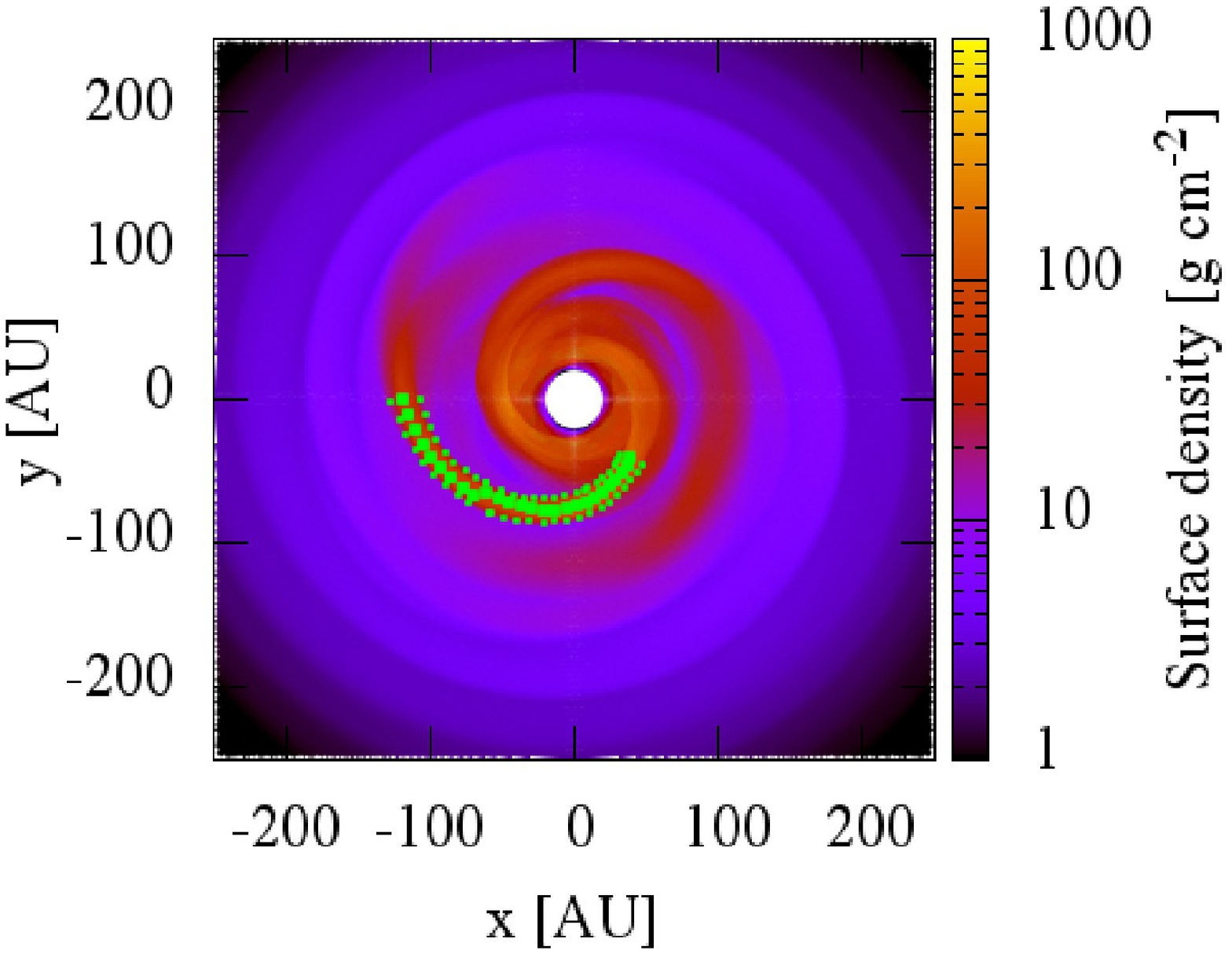}
\includegraphics[width=80mm]{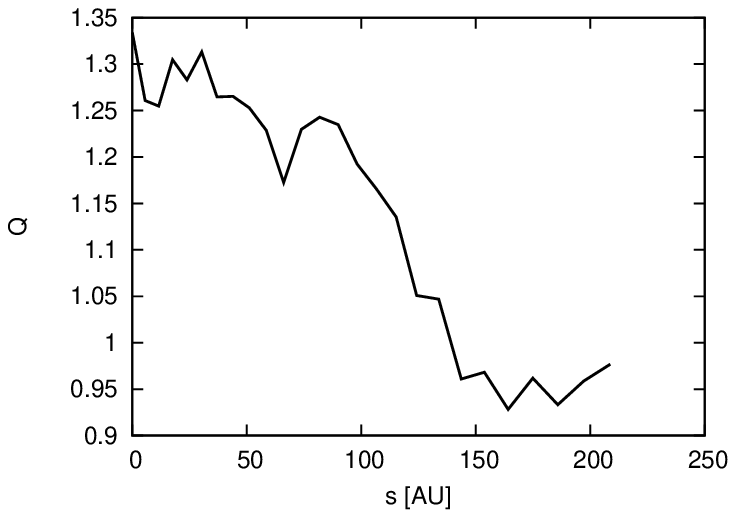}
\includegraphics[width=80mm]{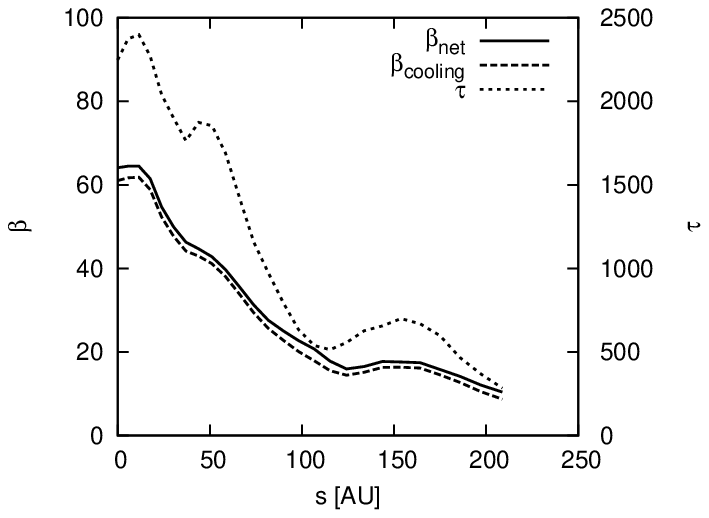}
\caption{Structure of the spiral arm at $t$=8428 years for model S265k05. This spiral arm  does not fragment.
  Thus, this result is also consistent with the criterion for the fragmentation: $Q <0.6$.}
\label{fig:spiral_large_kappa}
\end{figure}

The disk cooling rate depends on the opacity.
Since most of the regions of the disks are optically thick, the cooling rate is proportional to $\kappa_{\rm R}^{-1}$. 
Thus, the radiation cooling is efficient when $\kappa_{\rm R}$ is small.
To investigate the dependence of the condition of fragmentation on the cooling rate, 
we perform the simulations with $\kappa_{10}$ and $\Sigma_{100}$ different from Model S265k005. 
Additional models of parameters are listed in Table \ref{tab:models2}.
In models  S265k05, S265k025, S265k01, and S265k001, the initial surface density is as same as that of S265k005, but the opacity at $T=10$ K, $\kappa_{10}$, is 0.5, 0.25, 0.1, and 0.01 $[\rm cm^2 g^{-1}]$ respectively.
The results of model S265k005, S265k05, S265k025, S265k01, and S265k001, suggest that the fragmentation occurs in the case that the opacity is small and the radiation cooling is efficient if the other parameters are the same.
Fig. \ref{fig:spiral_large_kappa} shows the structures of the spiral arm in model S265k05, which does not fragment.
In the spiral arm, the minimum value of $Q$ is larger than 0.6 and 
the fact that this spiral arm does not fragment is consistent with our condition.
In this case, the opacity is an order of magnitude larger than the opacity of model S265k005,
and the normalized cooling time is $\beta_{\rm cool}\gtrsim 10$.
In the case that the opacity is large, the spiral arms cannot shrink enough to satisfy $Q\lesssim0.6$. 
In this sense the efficient cooling is important for the fragmentation of the spiral arms.

\begin{figure}
 \includegraphics[width=8cm]{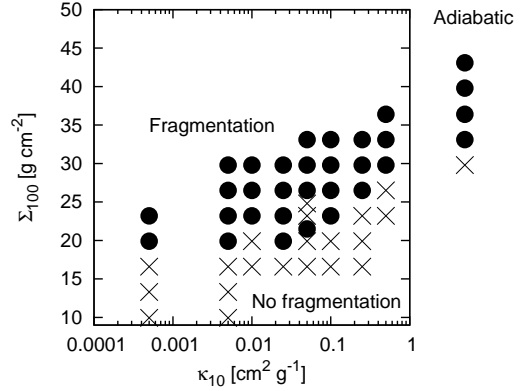}
\caption{Classification of the simulation results on the $\kappa_{10}-\Sigma_{100}$ plane. 
Filled circles and crosses denote fragmentation and no-fragmentation respectively. 
We also plot the results of the adiabatic calculations outside the square.}
\label{fig:sigma_kappa}
\end{figure}

\begin{figure}
 \includegraphics[width=8cm]{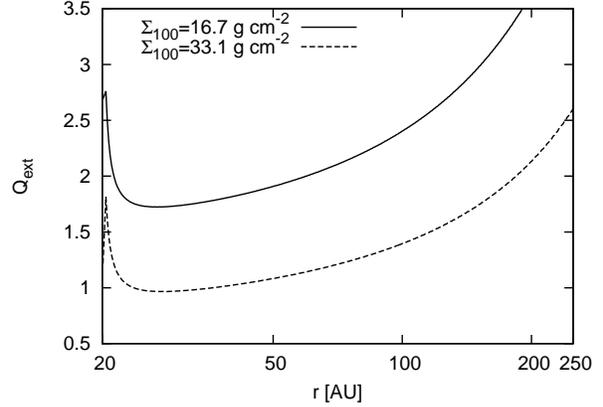}
 \caption{Distributions of $Q_{\rm ext}$ for $\Sigma_{100}=16.7$ and 33.1 $[\rm g \ cm^{-2}]$. 
In the case that $\Sigma_{100} \lesssim 16.7$  $[\rm g \ cm^{-2}]$, disks do not fragment independently of the opacity.  
In the case that $\Sigma_{100} \gtrsim 33.1$  $[\rm g \ cm^{-2}]$, disks fragment even in the adiabatic disks.}
 \label{fig:Qext}
\end{figure}

Fig. \ref{fig:sigma_kappa} shows the classification of the simulation results on the $\kappa_{10}-\Sigma_{100}$ plane. 
We also plot the result of the adiabatic calculations in which radiation cooling is shut off. 
The initial temperature of the adiabatic calculation is the same as $T_{\rm ext}$.
In the case that the opacity is small, the fragmentation occurs for the small initial surface density since the optical depth in the spiral arms are larger than unity and the cooling rate increases with decreasing the opacity. 
The critical value saturates at $\Sigma_{100}\sim 20\  [\rm g \ cm ^{-2}]$ in low opacity limit.
 The fragmentation occurs even in the adiabatic simulations when $\Sigma _{100} \gtrsim 30 \ [\rm g \ cm^{-2}]$.
Fig. \ref{fig:Qext} shows the radial distributions of $Q_{\rm ext}$ for  $\Sigma_{100}=16.7 $ and 33.1 $[\rm g \ cm^{-2}]$.
In the case that $\Sigma_{100}\lesssim 16.7 \ [\rm g \ cm^{-2}]$, $Q_{\rm ext}$ is larger than about 2, and the amplitude of the surface density of the spiral arms is too small to fragment independently of the opacity. 
In the case that $\Sigma_{100} \gtrsim 33.1 \ [\rm g \ cm^{-2}]$, $Q\sim1$ is satisfied and disks fragment independently of the opacity.


\subsection{Summary of counter-examples for cooling criterion}
\label{counter-examples}
\begin{figure*}
 \begin{minipage}{0.3\hsize}
  \begin{center}
   \includegraphics[width=60mm]{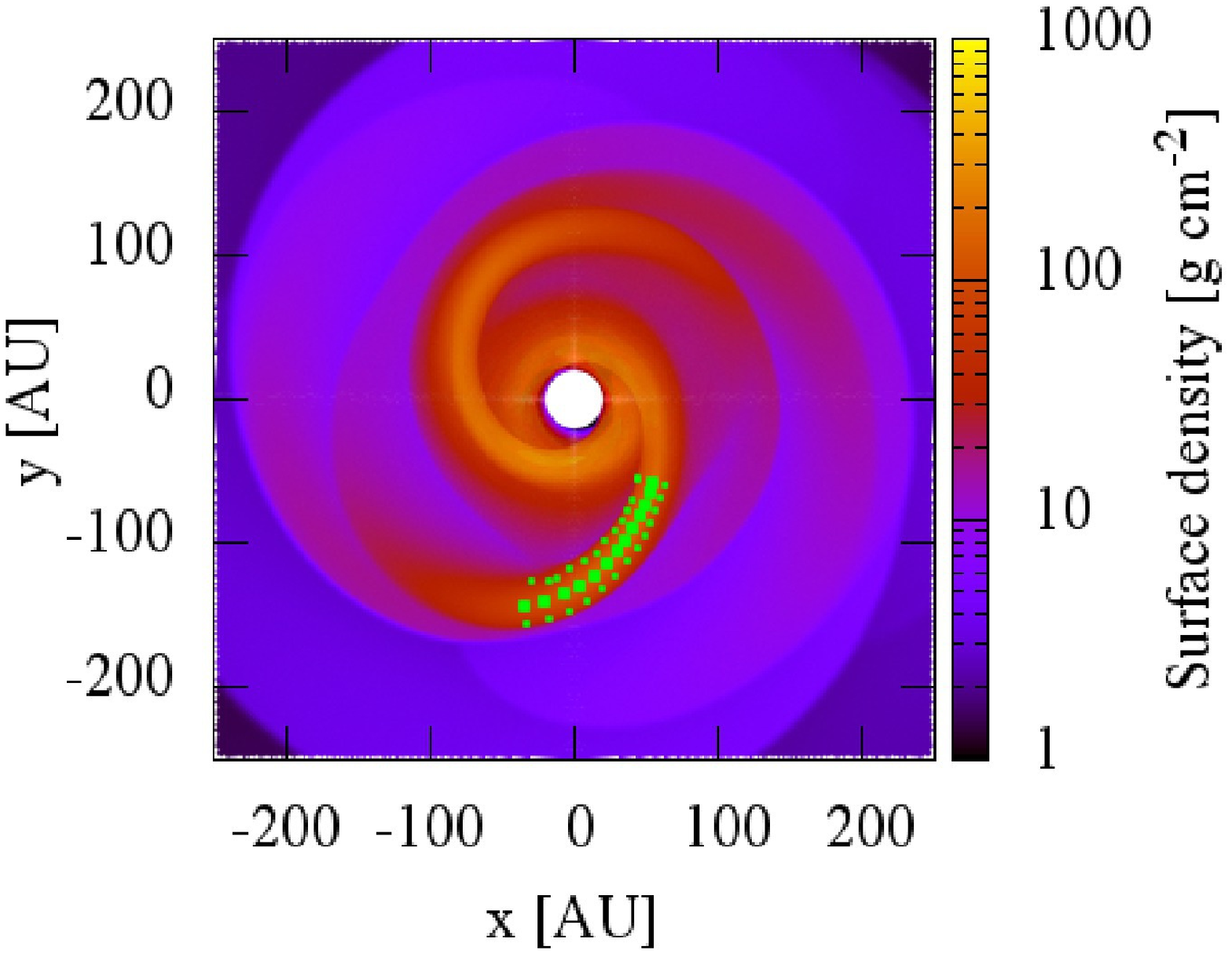}
  \end{center}
 \end{minipage}
 \begin{minipage}{0.3\hsize}
  \begin{center}
\includegraphics[width=60mm]{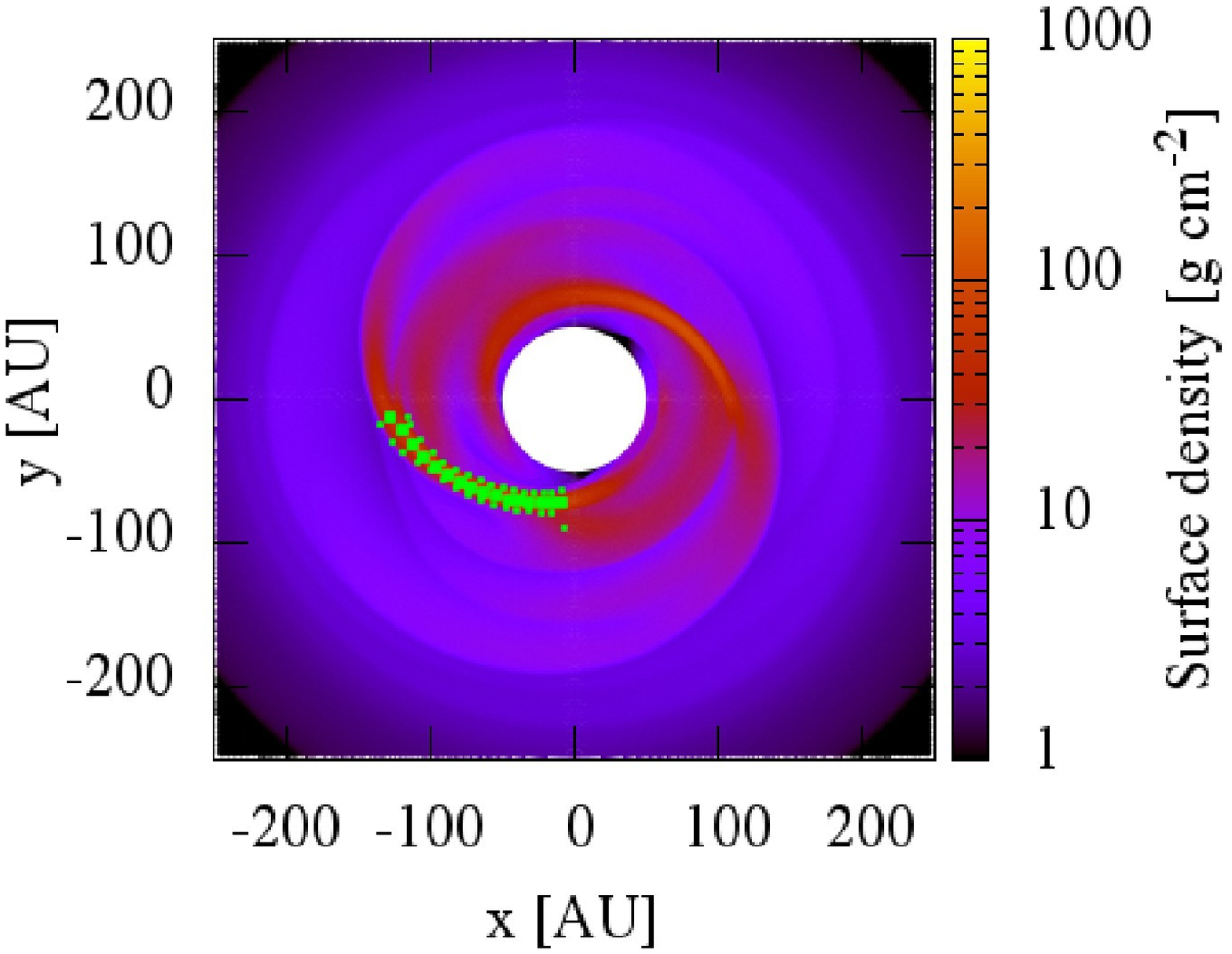}
  \end{center}
 \end{minipage}
 \begin{minipage}{0.3\hsize}
  \begin{center}
\includegraphics[width=60mm]{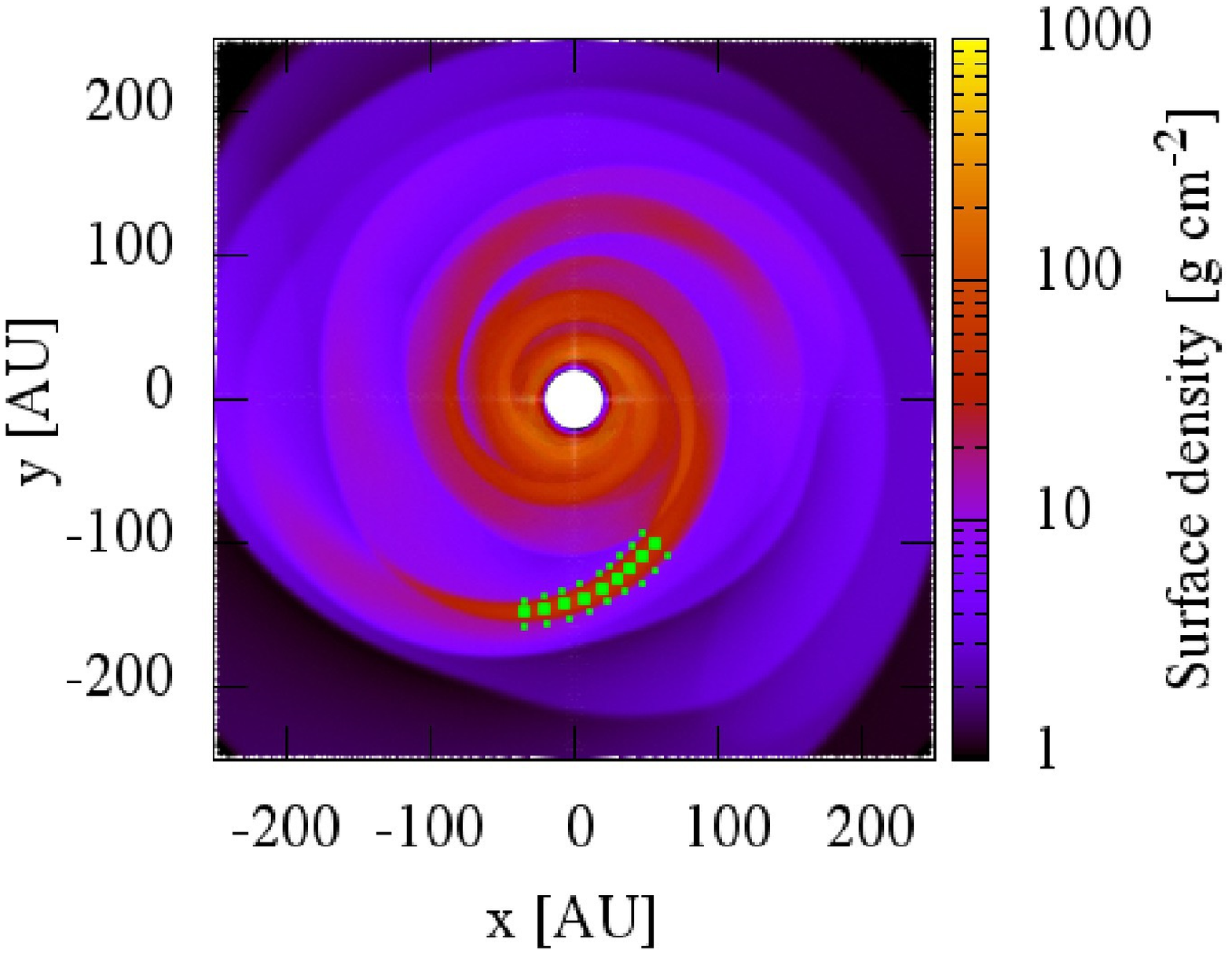}
  \end{center}
 \end{minipage}
 \begin{minipage}{0.3\hsize}
   \begin{center}
    \includegraphics[width=50mm]{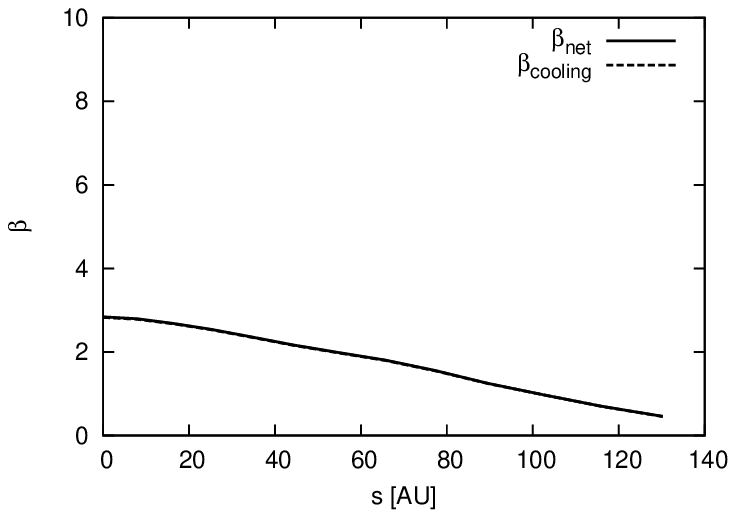}
   \end{center}
 \end{minipage}
 \begin{minipage}{0.3\hsize}
  \begin{center}
\includegraphics[width=50mm]{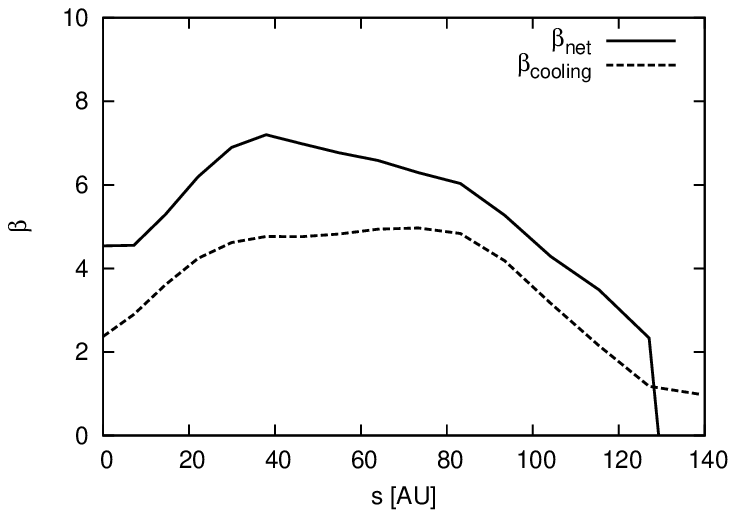}
  \end{center}
 \end{minipage}
 \begin{minipage}{0.3\hsize}
   \begin{center}
    \includegraphics[width=50mm]{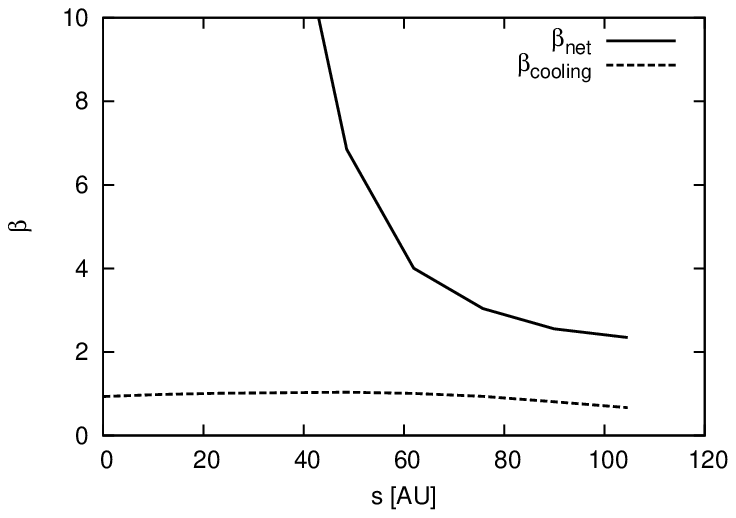}
   \end{center}
 \end{minipage}
\caption{Counter examples for the cooling criterion obtained by our simulations. 
The top panels show the surface density of the simulations and the bottom panels show $\beta_{\rm net}$ and $\beta_{\rm cool}$ in the spiral arms shown by green points in top panels. The models are S563Tx4 (left), Rin50 (middle), S265sft003(right).}
\label{fig:counter_not_frag}
\end{figure*}

In our simulations, we find many counter examples for cooling criterion. 
In this subsection, we summarize the results that conflict with the cooling criterion.
Fig \ref{fig:counter_not_frag} shows the results that the spiral arms do not fragment although they satisfy the cooling criterion.
In Fig. \ref{fig:counter_not_frag}, the top panels show the surface density of the simulations and the bottom panels show $\beta_{\rm net}$ and $\beta_{\rm cool}$ in the spiral arms shown by green points in the top panels. 
The models are S563Tx4 (left), Rin50 (middle), S265sft003(right).
In the spiral arms, $\beta_{\rm net}$ and $\beta_{\rm cool}$ are order unity, but the spiral arms do not fragment.
Especially, as described in Section \ref{largeT}, the result of S563Tx4 conflicts with the relation between $\beta_{\rm crit}$ and external irradiation obtained from local simulation done by \cite{2011MNRAS.418.1356R}.

Moreover, all the results of adiabatic simulation in which the fragmentation occurs are the counter examples for the cooling criterion.
These spiral arms fragment although the cooling time is infinitely large.
These results strongly suggest that the cooling criterion is neither a necessary condition nor a sufficient condition for fragmentation of massive irradiated protoplanetary disks. 
Thus we need a different criterion for gravitational fragmentation.
The criterion for the fragmentation of spiral arms obtained in this work may provide an important clue for finding more useful criterion for fragmentation of protoplanetary disks.

\section{Analytical consideration}
\label{linear_ana}
In Section \ref{result}, we proposed a new criterion for fragmentation of spiral arms: spiral arms fragment when $Q$ parameter inside the spiral arms becomes $Q<0.6$.
In this section, we perform the local linear stability analysis and show the criterion is also obtained analytically. 
Since the pitch angles of fragmenting spiral arms are small, we approximate spiral arms as rotating rigs around central stars.
We also assume that the wavenumber $k$ is small compared with the radius of the ring and  ignore the curvature of the ring.
Then the linear stability analysis of the ring is similar to that of the filament except for the effect of rotation. 
We adopt the local radial and azimuthal coordinate $(x,y)$.
We take into account the perturbations proportional to  $\exp[iky-i\omega t]$ and we use the linearised equations as follows:
\begin{equation}
  -i\omega \delta \Ml -ik\Ml \delta v_y = 0,
\end{equation}
\begin{equation}
 -i\omega \delta v_x = 2\Omega \delta v_y,
\end{equation}
\begin{equation}
 -i\omega \delta v_y = -2\Omega \delta v_x - ik\frac{\cs^2}{\Ml}\delta \Ml 
 -ik\delta\Phi,
\end{equation} 
where $\delta \Ml$ is the line mass perturbation of the ring, $\delta v_x$ and $\delta v_y$ are the perturbed velocities, $\Omega$ is the angular velocity of the background ring, and $\delta \Phi$ is the perturbed gravitational potential.
A rigid rotation is adopted since rotation profiles in the spiral arms observed in our numerical simulations are roughly approximated by rigid rotation because of their self-gravity.
Since we use the two-dimensional code in this work, we perform stability analysis of the flattened ring.
We evaluate the perturbed gravitational potential of the flattened ring by the summation of perturbed gravitational potential of the infinitesimally narrow filaments.
We approximately evaluate the perturbed gravitational potential of the filaments by using that of incompressible filaments.
This approximation enables us to obtain perturbed gravitational potential analytically.
Perturbed gravitational potential of the infinitesimally narrow filament is 
\begin{equation}
 d(\delta \Phi) = -2GK_0(kx)d(\delta M_{\rm L})
\end{equation}
where, $x$ is the distance from the axis of the filament, $d(\delta M_{\rm L})$ is the perturbed line mass of the infinitesimally narrow filament, $d(\delta \Phi)$ is the perturbed gravitational potential, and $K_0$ is the modified Bessel function of the second kind \cite[]{1961hhs..book.....C}. 
Then the perturbed gravitational potential of the flattened ring is given by the integration of the perturbed potential of the filament whose line mass is $d(\delta M_{\rm L}) = \delta \Sigma(x) dx$;
\begin{equation}
 \delta \Phi  =\int d (\delta \Phi)
  =\int -2G\delta \Sigma(x) K_0(kx) dx.
\end{equation}
For simplicity, we assume 
\begin{equation}
 \delta\Sigma(x) = \begin{cases}
			   \delta \Ml/(2W) &(-W <x< W ) \\
			   0 &(x<-W \ {\rm or}\  x>W ),
			  \end{cases}
\end{equation}
where $2W$ is the width of the flattened ring.
The gravitational potential at $x=0$ is
\begin{align}
 \delta \Phi & = \int_{-W }^{W } -G \delta \Ml K_0(|kx|)/W dx
  \nonumber \\
                & = -\pi G\delta\Ml
                [K_0(kW )L_{-1}(kW )\nonumber \\
 &\;\;\;\;\;\;\;\;\;\;\;\;\;\;\;\;\;\;\;\;
 +K_1(kW )L_0(kW )],
\end{align}
where $L_0$ and $L_{-1}$ are modified Struve functions.
In this case, we obtain the dispersion relation as follows;
\begin{align}
\omega^2 =&\cs^2k^2 -\pi G \Ml
  [K_0(kW)L_{-1}(kW)+K_1(kW)L_0(kW)]k^2 \nonumber\\
&+4\Omega^2.
\end{align}
We define the normalized frequency and wavenumber as follows;
\begin{equation}
 {\tilde \omega} = \frac{2W}{\cs}\omega,
\end{equation} 
\begin{equation}
 {\tilde k} = 2kW.
\end{equation}
The normalized dispersion relation is given by
\begin{align}
  {\tilde\omega}^2 = &{\tilde k}^2 -\frac{\pi G \Ml}{\cs^2}
  [K_0({\tilde k}/2)L_{-1}({\tilde k}/2)+K_1({\tilde k}/2)L_0({\tilde
 k}/2)]{\tilde k}^2 \nonumber \\
 &+\frac{16W^2\Omega^2}{\cs^2}.
\end{align}
We define two parameters, the normalized line mass and the ratio of the filament width and the scale height of the disk as follows:
\begin{equation}
 f \equiv \frac{G\Ml}{\cs^2},\label{eq:f}
\end{equation}
\begin{equation}
 l \equiv \frac{4W\Omega}{\cs}.\label{eq:l}
\end{equation}
We can rewrite normalized dispersion relation;
\begin{equation}
   {\tilde\omega}^2 = {\tilde k}^2 -  \pi f
  [K_0({\tilde k}/2)L_{-1}({\tilde k}/2)+K_1({\tilde k}/2)L_0({\tilde k}/2)]{\tilde k}^2 +l^2.
\end{equation}

\begin{figure}
 \centering
 \includegraphics[width=80mm]{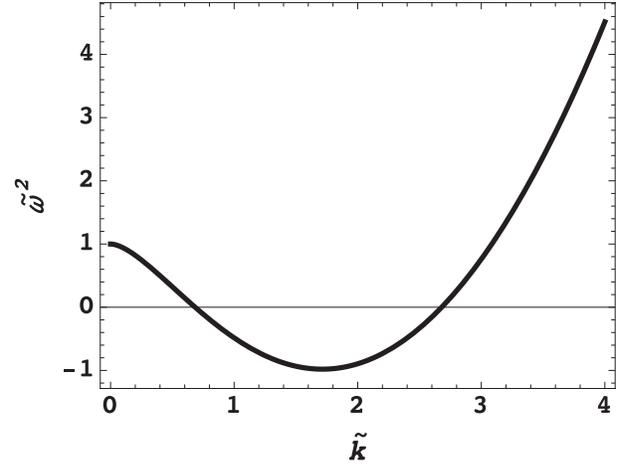}
 \caption{Dispersion relation of the flattened ring with $f=l=1$.
 The ring is unstable in the case that ${\tilde \omega}^2 <0$.}
 \label{fig:disp_kishimen_f1l1}
\end{figure}

\begin{figure}
 \centering
 \includegraphics[width=80mm]{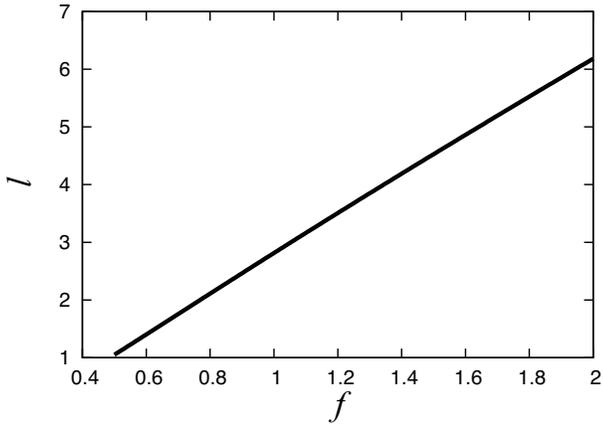}
 \caption{The relation between $f$ and $l$ that corresponds to marginally stable rings (${\tilde \omega}_{\rm min} =0$) .
 }
 \label{fig:f-l}
\end{figure}

\begin{figure}
 \centering
 \includegraphics[width=80mm]{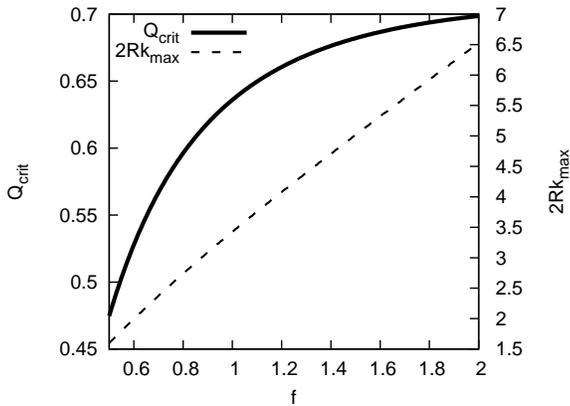}
 \caption{The critical $Q$ parameter and normalized most unstable wavelength $2Wk_{\rm max}$ as functions of $f$.
}
 \label{fig:f-Q_crit}
\end{figure}

\begin{figure}
 \centering
 \includegraphics[width=80mm]{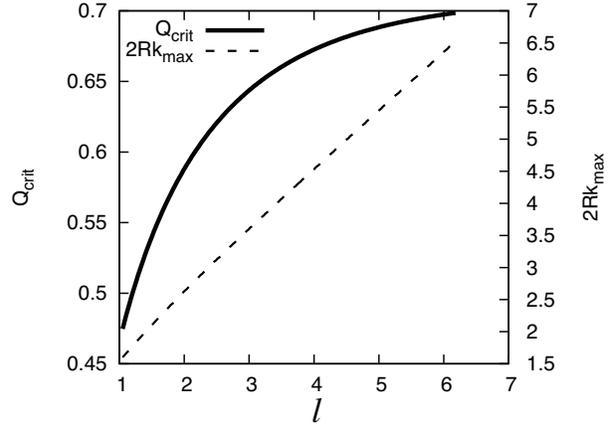}
 \caption{The critical $Q$ parameter and normalized most unstable wavelength $2Wk_{\rm max}$ as functions of $l$.
}
 \label{fig:l-Q_crit}
\end{figure}

Fig. \ref{fig:disp_kishimen_f1l1} shows the dispersion relation with $f=l=1$.
The flattened ring is unstable in the case that ${\tilde \omega}^2 <0$.
In the case that minimum value of ${\tilde \omega}^2$ is zero, we obtain the relation between $f$ and $l$ (Fig. \ref{fig:f-l}).
To compare the linear stability analysis with our numerical simulations, we need the relation of $Q$, $f$, and $l$.
The structures of the surface density of the spiral arms observed in our simulations are well fitted by a Gaussian function ($\Sigma(x) \propto \exp[-x^2/(2\sigma^2)]$).
Since the edges of the spiral arms are defined as the points where the surface density is 0.3 times the surface density at the center of spiral arms, the line mass of the spiral arms are given as follows:
\begin{equation}
 M_{\rm L} = \int_{-W}^W \Sigma(x)dx \approx 1.4 W \Sigma_{\rm max},
\end{equation}
where $2W$ is the width of the spiral arm and $\Sigma_{\rm max}$ is the surface density at the center of the spiral arm.
Thus, Toomre $Q$ parameter at the center of the spiral arms is 
\begin{equation}
 Q = \frac{2\cs\Omega}{\pi G \Sigma_{\rm max}} \approx  \frac{0.2l}{ f }.\label{eq:Q_fl}
\end{equation}
We can rewrite this relation by using $Q$ parameter.
The critical $Q$ for the instability is not unique but depends on $f$ or $l$.
The critical $Q$ and most unstable wavenumber as functions of $f$ are shown in Fig. \ref{fig:f-Q_crit}.
The relation between $Q_{\rm crit}$ and $f$ is fitted by
\begin{equation}
 Q_{\rm crit} \approx -0.11f^{-1.3}+0.744.
\end{equation}
The critical $Q$ and most unstable wavenumber as functions of $l$ are shown in Fig. \ref{fig:l-Q_crit}.
 Fig. \ref{fig:f-Q_crit} shows that in the case that $f \sim 1$, the critical $Q$ parameter for the instability is about $0.6$. 
This result is consistent with the results of the numerical simulation performed in this work.
This result shows that the condition of gravitational fragmentation of the protoplanetary disks is given by the condition of the gravitational instability of the spiral arms formed in the disks.
The $Q$ parameter used in this analysis is defined by local values of $\Sigma$, $\cs$, and $\Omega$.
When we apply this criterion to realistic systems, we should use the minimum $Q$ parameter in the spiral arm.

\section{Discussion}
\label{discussion_frag}
\subsection{Comparison with previous work}
The normalized cooling timescale has been considered to be a critical parameter for the fragmentation of the protoplanetary disks \cite[]{2001ApJ...553..174G,2005MNRAS.364L..56R,2005ApJ...621L..69R,2011ApJ...740....1K}.
As shown in this paper, however, the cooling timescale cannot be used as a useful criterion to predict the stability of the disk against self-gravitational fragmentation.
In this section, we compare our results with the results of previous work.

The discrepancy may come from the difference of the disk masses.
Since previous numerical simulations on protoplanetary disk formation suggest that the initial disk mass is comparable to or larger than the central star mass \cite[]{2010ApJ...718L..58I,2010ApJ...724.1006M,2011ApJ...729...42M,2011PASJ...63..555M,2014MNRAS.438.2278M,2012PTEP.2012aA307I}, we adopt the massive disk as initial conditions (for example, initial disk mass of S265k005 is $\approx 0.38 M_*$).
\cite{2005MNRAS.358.1489L} have pointed out that the nature of the spiral arm in the massive disk is different from that of less massive disk.
The scale height of a marginally unstable disk ($Q \sim 1$) increases  as the disk mass increases, because the temperature of the disk increases when $Q$ is fixed and $\Sigma$ increases \cite[cf.][]{1999A&A...350..694B}.
As a result, grand design spiral arms with wavenumber $m\sim 2$ are formed in a massive disk.
On the other hand, in the previous studies on the cooling criterion, tiny spiral arms with large $m$ are formed in the disks because the disk mass is much smaller than the central star mass.
Since the disk mass affects the nature of the spiral arms formed in the disks, the difference in the disk mass may affect the property of fragmentation.
Moreover, \cite{2010ApJ...708.1585K} have performed the three-dimensional numerical simulation of disk formation and pointed out that the condition for fragmentation of the disk depends on the disk mass.
These previous studies have suggested that the condition for fragmentation depends on the disk mass and it will be one of the reasons why the cooling criterion cannot predict the fragmentation of the disks observed in this work.

Another reason for the discrepancy is in the modeling of radiative transfer.
Many previous calculations on the cooling criterion modeled radiative transfer with cooling function and neglected the external radiation \cite[]{2005MNRAS.364L..56R, 2011MNRAS.416L..65P, 2012MNRAS.420.1640R, 2014MNRAS.438.1593R}.
With this simplification, temperatures of spiral arms continue to decrease unless heating by the collision of the spiral arms balances the cooling.
Consequently, the cooling timescale determines disk fragmentation.
On the other hand, in realistic situations, stellar irradiation provides the dominant energy source except for inner region (see Fig. \ref{fig:rad_ave_long_disk_2}).
Moreover, the disks can be heated by the radiation from the inner disks even when the radiation from the central star is neglected \cite[]{2015MNRAS.446.1175T}. 
Thus the external irradiation cannot be neglected in the investigation of a realistic criterion of gravitational fragmentation of disks.
In realistic case, the temperatures of the spiral arms are roughly given by the balance between the external irradiation and the radiative cooling,
and the thermal pressure can support the spiral arms without heating by gravitational instability.
Therefore, $\beta$ cannot predict the disk fragmentation in the irradiated disks.

Our results do not conflict with previous work that investigates fragmentation of the irradiated disk 
\cite[e.g.][]{2008A&A...480..879S,2009MNRAS.392..413S}.
\cite{2008A&A...480..879S} have shown a certain set of non-fragmenting disks that do not satisfy the cooling criterion, and  \cite{2009MNRAS.392..413S} have shown a certain set of fragmenting disks that satisfy the cooling criterion.
Indeed, if we choose the disks that are plotted near the boundary between fragmentation and non-fragmentation in Fig. \ref{fig:sigma_kappa}, they might appear virtually consistent with the cooling criterion. 
However, this does not mean that the cooling criterion is a meaningful criterion. We can find many counter-examples for the cooling criterion. 
In other words, the previous work only considered limited parameter space that does not include cases inconsistent with the cooling criterion. 
In this way, we can understand the difference in the conclusions. 

There is other previous work that is consistent with our conclusion.
\cite{2015MNRAS.446.1175T} showed an example that the disks  do not fragment even when the disks satisfy the condition for the cooling rate.
On the other hand, the fragmentation occurs even in the case where almost adiabatic equation of state is adopted \cite[e.g.][]{2010ApJ...724.1006M, 2011ApJ...729...42M}.
Thus the condition that the cooling timescale is smaller than the critical value is neither sufficient nor necessary conditions for the fragmentation of the disk. 

\cite{2015MNRAS.446.1175T} suggest that the critical $Q$ parameter for the fragmentation is $Q\sim 0.3$ and smaller than our result, 0.6. This difference would come from the definition of $Q$: in \cite{2015MNRAS.446.1175T}, angular velocity $\Omega$ is used to calculate $Q$ instead of $\Omega_{\rm epi}$.
Our numerical experiments show that in self-gravitating disk, $\Omega$ and $\Omega_{\rm epi}$ can be largely different.

\cite{2009MNRAS.393.1157C} investigate the relation between  
$\beta$
and surface density perturbations $\delta \Sigma \equiv \Sigma - \overline{\Sigma}$,
where $\overline{\Sigma}$ is azimuthal averaged surface density.
The relation has been obtained from global three-dimensional simulations;
\begin{equation}
 \frac{\delta \Sigma}{\overline{\Sigma}} \sim \frac{1}{\sqrt{\beta}}.
\end{equation}
\cite{2009MNRAS.393.1157C} have also mentioned $\beta_{\rm crit} \sim 4$ meaning that the 
critical surface density perturbation is $\delta \Sigma \sim 0.5\Sigma$.
This result may be understood by our criterion for fragmentation of spiral arms.
The quasi-steady disk satisfies the azimuthally averaged $Q$ parameter $\sim 1$.
If the difference between $Q$ in the spiral
arm and averaged $Q$ parameter comes from the surface density perturbation
(this means that the rotation profile and temperature in the spiral arm
are almost the same as the averaged value),
the $Q$ in the spiral arm is given as
\begin{equation}
Q_{\rm spiral} =
 \frac{\cs\Omega_{\rm epi}}{\pi G (\overline{\Sigma}+\delta \Sigma)}
\approx
  0.67
\end{equation}
This value is similar to the critical $Q$ value obtained in this work.
This suggests that the condition for fragmentation of spiral arms found in this work
can also explain the previous studies that use simple cooling function and neglect external irradiation.

\cite{2012MNRAS.421.3286P} performed the two-dimensional, local numerical simulations similar to the simulation performed by \cite{2001ApJ...553..174G}, but the integration time is longer than  \cite{2001ApJ...553..174G}.
They found that there is no sharp boundary in the cooling time between disks in which fragmentation occurs and disks in which fragmentation does not occur. 
This is because fragmentation is a nonlinear outcome of gravitational instability and depends sensitively on an initial condition.
Such a nature of the fragmentation also appears in our results.
For example, the fragmentation occurs in model S215k005 and does not occur in model S232k005, although initial surface density of model S215k005 is smaller than that of model S232k005.
We conjecture that the actual boundary of fragmentation and non-fragmentation in Fig. \ref{fig:sigma_kappa} is not simple but possibly fractal in the limit of long timescale evolution. 

In many previous calculations on the cooing criterion, they fixed an initial surface density and changed the normalized cooling time $\beta$.
They correspond to our calculations that the initial surface density is fixed and opacity is changed in Fig. \ref{fig:sigma_kappa}.
Fig. \ref{fig:sigma_kappa} shows that if the opacity is small and the radiative cooling is efficient, the spiral arms easily satisfy the condition for the fragmentation proposed in this work.
However, it is difficult to obtain the critical cooling timescale quantitatively from such calculations since the criterion for the gravitational fragmentation depends strongly on the surface density but weakly on the cooling timescale, which is visible as a gentle slope of the boundary between fragmentation and non-fragmentation in Fig. \ref{fig:sigma_kappa}.

\begin{figure}
\begin{center} 
 \includegraphics[width=8cm]{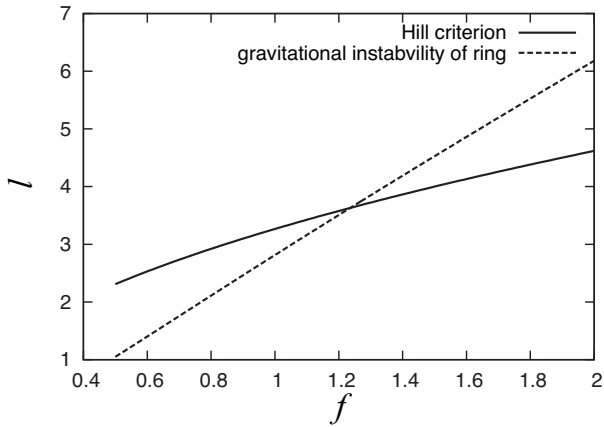}
\end{center}
\caption{Critical $l$ for the fragmentation of the spiral arms as a function of $f$.
The $l$ smaller than the solid line satisfied Equation (\ref{eq:Hill_criterion})
and the dashed line shows the criterion of the gravitational instability of the spiral arms shown in Fig. \ref{fig:f-l}.}
\label{fig:Hill_crit}
\end{figure}

\begin{figure}
\begin{center} 
 \includegraphics[width=8cm]{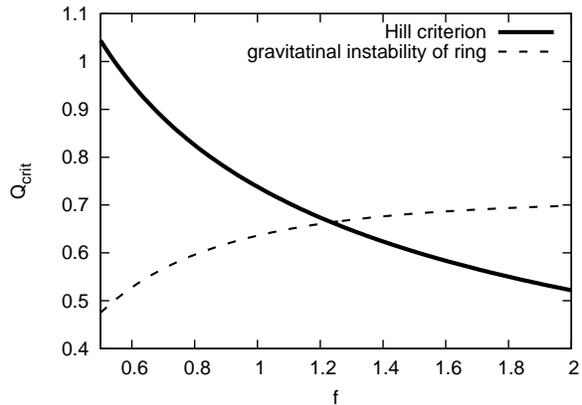}
\end{center}
\caption{Critical $Q$ parameter for the fragmentation of the spiral arms as a  function of $f$.
The solid line shows the criterion given by Hill criterion (Equation (\ref{eq:Hill_Q_crit}) ) and the dashed line shows the criterion of the gravitational instability of the spiral arms shown in Fig. \ref{fig:f-Q_crit}.}
\label{fig:Hill_Q_crit}
\end{figure}

\cite{2012MNRAS.423.1896R} have proposed ``Hill criterion'' for fragmentation of spiral arms.
They approximate that the mass of the fragment formed in the spiral arm is $2RM_{\rm L}$, where $M_{\rm L}$ is the line mass and $2R$ is the width of spiral arms.
They claimed that the spiral arms fragments when the width of the spiral arm becomes smaller than the Hill radius given by the mass $2RM_{\rm L}$;
\begin{equation}
2R< 2H_{\rm Hill} =
2\left(\frac{2GM_{\rm L}R}{3\Omega^2}\right)^{1/3}.
\label{eq:Hill}
\end{equation}
We can rewrite Equation (\ref{eq:Hill}) by using the parameters $f$ and $l$ (Equation (\ref{eq:f}) and (\ref{eq:l})) as follows:
\begin{equation}
 l < \sqrt{\frac{32}{3}f}.
\label{eq:Hill_criterion}
\end{equation}
Fig. \ref{fig:Hill_crit} shows criterion of the fragmentation given by Equation (\ref{eq:Hill_criterion}). 
We also plot the criterion of the gravitational instability of the spiral arms shown in Fig. \ref{fig:f-l}.
We obtain the critical $Q$ parameter as a function of $f$ by using Equation (\ref{eq:Q_fl}) and (\ref{eq:Hill_criterion}) as follows:
\begin{equation}
 Q \lesssim 0.7 f^{-1/2}.
\label{eq:Hill_Q_crit}
\end{equation}
Fig. \ref{fig:Hill_Q_crit} shows the relation between the critical $Q$ parameter and $f$ obtained by Equation \ref{eq:Hill_Q_crit}.
We also show the critical $Q$ parameter of the gravitational instability of rings.
Both criteria give similar results around $f\sim 1.2$.
Since $f\sim 1$ is realized in fragmenting spiral arms in our simulations, our numerical results seem consistent with \cite{2012MNRAS.423.1896R}.
However, there is a remarkable difference of these two critical $Q$ parameters in the case that $f$ is small.
For fixed $Q$ parameter, spiral arms are gravitationally unstable if $f$ is large since large $f$ means strong gravitational force. 
On the other hand, spiral arms satisfy Hill criterion if $f$ is small since small $f$ for fixed $Q$ means small spiral width.
However, it seems difficult that fragmentation occurs in such spiral arms since they are gravitationally stable.

\subsection{Formation of spiral arms that satisfy the condition of
  fragmentation}
This work reveals that $Q$ parameter in the spiral arm is essential for gravitational fragmentation of protoplanetary disks: fragmentation occurs when $Q<0.6$ in the spiral arm.
Thus we can divide the process of fragmentation into two stages: formation of spiral arms due to global gravitational instability of protoplanetary disks; and fragmentation of spiral arms due to gravitational instability of the spiral arms.
Our work gives the criterion for the fragmentation in the second stage.
Therefore, our work reduces the condition for the fragmentation of the disks to the condition for formation of spiral arms that satisfy $Q < 0.6$.

There are some variations in the formation processes of the spiral arms that satisfy the condition for fragmentation $Q<0.6$.
In the case that disks are massive enough and $Q$ is small, two spiral arms are formed in a dynamical timescale.
The outer parts of the spiral arms are tightly wound and they collide each other.
By this collision, the surface densities become large enough to satisfy the condition $Q<0.6$ around the collided region.
In the case that disks are less massive, the formation process of the spiral arms that satisfy the condition $Q<0.6$ takes a long time and is more complicated.
For example, we show the fragmentation process of the spiral arm of model S265k005 shown in Fig. \ref{fig:spiral_fragment}.
This spiral arm collide another one at about 4000 yr and the high-surface density region is formed in the spiral arm.
However, the condition $Q<0.6$ is not satisfied in this region and the fragmentation does not occur immediately. 
The spiral arm propagates outward  ($v_{r}>0$ in the most of the region in this spiral arm) and the high-surface density region also moves outward.
After that, the region satisfies the condition $Q<0.6$ and then the spiral arm fragment.
In these cases, the collision and propagation of the spiral arms are important for the process of the fragmentation of the spiral arms.

After the fragmentation, spiral arms are formed through gravitational interaction between the fragments and the disk gas.
An additional fragmentation often occurs in such spiral arms.
In this case, the fragments formed initially accrete onto the central  stars and the fragments formed secondly often survive.  
Thus such a sequential fragmentation may be important for the formation process of the gas giant planets or binary systems.  

Previous theoretical work on the self-gravitating protoplanetary disks used the one-dimensional accretion disk model
\cite[cf.][]{2010ApJ...713.1143Z, 2013ApJ...770...71T, 2013ApJ...774...57B}.
To conveniently describe possible fragmentation process in such a simplified protoplanetary disk model, the criterion for the fragmentation given by the azimuthally averaged value is required.
To obtain the criterion, we need further analysis on the formation and propagation process of the spiral arms that satisfy $Q<0.6$.

\section{Conclusions}
\label{conclusion_frag}
In this work, we investigated the physical process of self-gravitational fragmentation of protoplanetary disks to clarify the reliable criterion of the fragmentation. 
We performed both two-dimensional numerical simulations of self-gravitating protoplanetary disk and linear stability analysis for gravitational instability of spiral arms.
We took into account external irradiation from a central star to obtain a realistic criterion.
We found that the fragmentation occurs when the spiral arms satisfy $Q \lesssim 0.6$.
We conducted numerous simulations by changing parameters of the opacity, the external radiation, the inner radius, and the softening length of the disks and confirmed that this criterion is always valid.
We have further confirmed that this fragmentation criterion can be obtained from the linear stability analysis of self-gravitating spiral arms.

Our results showed that the cooling criterion ($\Omega \tau_{\rm c} \lesssim O(1)$) is neither a necessary condition nor a sufficient condition for fragmentation of massive, irradiated protoplanetary disks.
We obtain many examples in which the fragmentation does not occur even if the cooling time is small enough. 
On the other hand, we found the cases in which the fragmentation occurs even if we adopt an adiabatic EOS (infinite cooling time).
In all cases, the results are consistent with our condition for fragmentation of spiral arms ($Q<0.6$).
Our results suggest that the $Q$ parameter in the spiral arms is more essential parameter than the normalized cooling time for fragmentation of spiral arms.
We can divide the process of fragmentation into two stages: the first stage corresponds to the formation of spiral arms and the second their fragmentation. 
Our work provides a clear criterion for the second stage: the condition for the fragmentation of the disks is given by the condition for formation of spiral arms that satisfy $Q < 0.6$.

\section*{Acknowledgments}
We thank Kazuyuki Omukai for his continuous encouragement and comments.
We also thank Kazuyuki Sugimura, Shigeo Kimura, and Sho Fujibayashi for fruitful discussion.
This work was partly supported by a Grant-in-Aid for JSPS Fellows and JSPS KAKENHI Grant Numbers 23244027 and 23103005.

\end{document}